\documentclass[fleqn,usenatbib]{mnras}

\usepackage{newtxtext,newtxmath}

\usepackage[T1]{fontenc}

\DeclareRobustCommand{\VAN}[3]{#2}
\let\VANthebibliography\thebibliography
\def\thebibliography{\DeclareRobustCommand{\VAN}[3]{##3}\VANthebibliography}

\usepackage{graphicx}	
\usepackage{amsmath}	
\usepackage{amsmath,amssymb}	
\usepackage{lscape}
\usepackage{threeparttable}
\usepackage{arydshln}
\usepackage{multirow}
\usepackage{chngpage}
\usepackage{pdfpages}
\usepackage{footmisc}

\title[Chemical evolution of dwarf galaxies]{The impact of rare events on the chemical enrichment in dwarf galaxies}

\author[]{
Nao Fukagawa,$^{1,2}$\thanks{E-mail: nao.fukagawa@nao.ac.jp, naofukagawa@gmail.com}
Nikos Prantzos$^{3}$
\\
$^{1}$The department of astronomical science, The Graduate University for Advanced Studies, SOKENDAI, 2-21-1 Osawa, Mitaka, Tokyo 181-8588, Japan\\
$^{2}$National Astronomical Observatory of Japan, 2-21-1 Osawa, Mitaka, Tokyo 181-8588, Japan \\
$^{3}$Institut d'Astrophysique de Paris, UMR7095 CNRS, Sorbonne Universit\'{e}, 98 bis Bd. Arago, 75014 Paris, France
}

\date{Accepted XXX. Received YYY; in original form ZZZ}

\pubyear{2023}

\begin{document}
\label{firstpage}
\pagerange{\pageref{firstpage}--\pageref{lastpage}}
\maketitle

\begin{abstract}
In the environments where the abundance of heavy elements is low, 
rare events are expected to impact the chemical enrichment. 
Dwarf galaxies have small masses, low average metallicities and in general low star formation rates, 
and thus investigating the chemical enrichment 
provides understanding on the impact of each source of elements 
on the chemical abundance. 
Using a chemical evolution model in which the rarity is introduced, 
we investigate the impact of rare events on the chemical enrichment 
for Local Group dwarf galaxies. 
In the model, the occurrence of individual sources of elements is 
estimated with the star formation history derived by the colour-magnitude 
diagram. 
The abundance ratios of trans-iron elements to iron predicted by the model 
show the oscillation at the lowest metallicities 
because of the r-process events. 
In the case of a galaxy of a lower mass, the oscillation caused 
by neutron star mergers is also seen at higher metallicities, 
which suggests that the rarity can be important in lower-mass systems.
Regarding the source of the chemical enrichment, 
we observe 
that the r-process sites seem to contribute more to 
the production of trans-iron elements at low metallicities,
but massive stars of different rotating velocities 
also contribute to create part of the dispersion of the abundance ratios
through the s-process.
Both observational and theoretical data, including nucleosynthesis calculations and the chemical abundance
of metal-poor stars, are needed to obtain deeper insights into the sources 
of the chemical enrichment at low metallicities.
\end{abstract}





\section{Introduction}

Investigating the chemical enrichment in low-metallicity systems is 
important to understand the sources of elements in 
the pristine environments. 
Stars formed from the metal-poor gas have been suggested to rotate rapidly,
release the products of nucleosynthesis to the interstellar medium and
contribute to the chemical enrichment.
In particular, massive stars have short lifetimes ($\sim10^7$ years), and 
rotating massive stars are supposed to contribute to the enrichment of nuclei lighter 
than the Fe-peak, such as ${\rm {}^{14}N}$ \citep[e.g.][]{2002A&A...390..561M,2007A&A...461..571H} and  ${\rm {}^{19}F}$ \citep{2018ApJS..237...13L}, and 
the trans-iron elements \citep[e.g.][]{2012A&A...538L...2F,2016MNRAS.456.1803F}
through stellar winds and supernovae
in the early phase of the evolution of systems 
\citep[e.g.][]{2006A&A...449L..27C,2013A&A...553A..51C,2018MNRAS.476.3432P}.

Regarding the sources of elements heavier than the Fe-peak, the neutron capture process, 
including the slow (s-) and the rapid (r-) 
processes, are responsible\footnote{Nuclei heavier than the Fe-peak are also 
proposed to be synthesized through other processes, 
such as the intermediate neutron capture process \citep[][]{1977ApJ...212..149C}, 
the p-process \citep[][]{1976A&A....46..117A} and 
the non-thermal nucleosynthesis \citep[][]{2022A&A...658A.197G}.} 
for the production. 
The s-process proceeds in the interior of stars on long time-scales 
($\sim 10^4$~years), 
and the r-process occurs on short time-scales 
($\sim 1$~second) in the explosive environment of a large neutron flux 
\citep{1957RvMP...29..547B}.  
The contribution of the s- and r- processes to the solar abundance has been 
assessed in previous studies 
\citep[e.g.][]{1999A&A...342..881G,2004ApJ...601..864T,2020MNRAS.491.1832P}.
Since the products of nucleosynthesis of stars vary with metallicity, 
the contribution of the s- and r- processes to the chemical enrichment 
at low metallicities may not always be compared with that 
in the solar neighbourhood.

The chemical abundance of stars reflects the chemical composition of gas 
that formed the stars. 
It is widely known that there is the dispersion in the ratios of the abundances
of trans-iron elements to that of iron of stars in the Milky Way at low metallicities which cannot be explained solely by the uncertainty about the measurement
\citep[e.g.][]{1988ApJ...327..298G,1991AJ....102..303R,1995AJ....109.2757M}.
The scatter is supposed to reflect physical processes, 
such as the nucleosynthesis processes
\citep[e.g.][]{2014ApJ...797..123H}, 
different conditions of supernovae \citep[e.g.][]{2017ApJ...836L..21N}, 
the inhomogeneous mixing in the halo and the stochastic and rare star formation 
\citep[e.g.][]{1999ApJ...511L..33I,2000A&A...356..873A,2008A&A...481..691C,2015MNRAS.452.1970W}
and the natal kick of neutron stars \citep{2022MNRAS.512.5258V}.
In the earliest phase of the chemical enrichment,
a small number of the first massive stars inhomogeneously enriches 
the surroundings in different elements,
which might create the dispersion in the abundance ratios of metal-poor stars
\citep[e.g.][]{1995ApJ...451L..49A}. 

In low-metallicity environments, 
individual rare events are expected to impact the chemical abundance 
more significantly than in the environments enriched in heavy elements. 
Importantly, according to the cold dark matter model 
\citep[e.g.][]{1978MNRAS.183..341W},  
more massive systems are formed through mergers of less massive systems.
Based on the correlation between the galaxy mass and the average metallicity 
\citep[e.g.][]{2004ApJ...613..898T,2013ApJ...779..102K},
galaxies of lower masses 
tend to have lower metallicities.
Therefore, the role of rare events on the chemical enrichment in low-mass and 
low-metallicity systems needs to be investigated.

Dwarf galaxies have small masses in stars 
(typically smaller than ${\rm 10^{8-9}M_{\odot}}$)
and low average metallicities\footnote{Hereafter, we refer to [Fe/H] as
metallicity. [A/B] represents the abundance ratio,
given by [A/B]${\rm \equiv \log(N_A/N_B)-\log(N_A/N_B)_{\odot}}$, where ${\rm N_A}$ denotes
the abundance of A.}.  
Also, in general, dwarf galaxies tend to have low average star formation rates. 
For instance, the maximum of the star formation rate throughout the 
evolution of Fornax dwarf spheroidal galaxy, 
one of massive Local Group dwarf galaxies 
(${\rm 2.5\times10^7~M_{\odot}}$)\footnote{The stellar mass listed in Table~4 
in \citet{2013ApJ...779..102K} is referred to. 
The value has been derived with the luminosity taken 
from \citet[and references therein]{2012AJ....144....4M} and the stellar mass-to-light ratio 
derived by \citet{2008MNRAS.390.1453W} for individual dwarf galaxies.\label{foot:mass}}
has been estimated to be about 
${\rm 3.5\times10^{-3}~M_{\odot}yr^{-1}}$
\citep[][]{2012A&A...544A..73D}, 
while the present-day star formation rate of our Galaxy is about ${\rm 1~M_{\odot}yr^{-1}}$
\citep[][]{2010ApJ...710L..11R}. 
When a star formation rate in a system is low, the number of stars formed 
from gas per unit time and that of explosive events that release heavy elements 
are expected to be small. 

To date, over 100 dwarf galaxies have been found in the Local Group
\citep[][for a review of Local Group dwarf galaxies]{2009ARA&A..47..371T}.
The chemical composition of individual stars in the nearby dwarf galaxies
that have smaller masses than the Magellanic Clouds 
has been measured. 
Also, the star formation history of individual dwarf galaxies has been 
derived based on the colour-magnitude diagram 
\citep[e.g.][]{1991AJ....102..951T,2002MNRAS.332...91D}.
Since abundance ratios depend on the star formation history, 
the observationally-derived star formation history can be a constraint on 
chemical evolution models and allows us to discuss the impact of each source 
of heavy elements on the chemical abundance.

Astrophysical sources of the chemical enrichment have been discussed
by comparing observational data with chemical evolution models where
the stochasticity is incorporated
\citep[e.g.][]{2015A&A...577A.139C,2018ApJ...865...87O};
see next section for a brief description of previous studies.
In this study, we take into account rare events in a chemical evolution model 
in a different way and attempt to discuss 
(i) the role of rare events on the chemical enrichment and 
(ii) the sources of heavy elements in the chemical enrichment for two 
dwarf spheroidal galaxies (dSphs) around the Milky Way.
Firstly, we explain our definition of rare events in this study
(Section~\ref{sec:2}).
The concept is adopted to the dwarf galaxies (Section~\ref{sec:3}).
With the numerical chemical evolution model for dwarf galaxies
that includes rare events in the source of the chemical enrichment,
we compare the observables (the metallicity distribution and 
the abundance ratio)
of the galaxies with those derived by the model and attempt to interpret 
the results (Section~\ref{sec:4}).
We also discuss the impact of individual events on the chemical enrichment
in ultra-faint dwarf galaxies (Appendix~\ref{sec:app-uFd})
and in a cosmological context (Appendix~\ref{sec:app-bb}, online material).



\section{Concept of rare events}
\label{sec:2}

One of interpretations of the dispersion in abundance ratios of 
the Galactic halo stars is the incomplete mixing of the interstellar gas
in the early phase of the formation.
This has been investigated with stochastic chemical evolution models
where the star formation is assumed to be induced by supernovae
\citep[e.g.][]{1999ApJ...511L..33I,2004A&A...416..997A} or
collisions of clouds \citep[][]{2004ApJ...601..864T}.
In addition to the incomplete mixing of gas, the variation in mass of ejecta
(yields) released by r-process events has been 
included in models \citep[e.g.][]{2015A&A...577A.139C}.

The hierarchical formation of the Galactic halo is also relevant to the 
stochasticity.
For instance, in \citet{2018ApJ...865...87O}, the Galactic halo is assumed to
be formed by mergers of sub-haloes with different star formation histories
and age -- metallicity relations.
At each time, a sub-halo is chosen and a neutron star merger (NSM) occurs there.
The abundance ratio of the sub-halo may be quite different from
others and that creates the dispersion in the abundance ratio at a given
metallicity.

In the present study, we focus on the impact of rare events in
low-mass galaxies based on their observationally-derived star formation
history and introduce the stochasticity to the occurrence of each
astrophysical source of the chemical enrichment.

We roughly estimate how many events of astrophysical sources of elements 
can occur in a dwarf galaxy of stellar mass ${\rm M_* = 10^6~M_{\odot}}$. 
We set the time-step ${\rm \Delta t}$ to be ${\rm \Delta t = 10^{-3}}$
Gyr, which is comparable to or shorter than a typical lifetime of massive stars.
If stars are constantly formed during ${\rm T_{SF} = 10}$~Gyrs, 
this dwarf galaxy has a star formation rate 
${\rm \Psi = M_* / T_{SF} = 10^6~M_{\odot} / 10~Gyr = 10^5~M_{\odot}~Gyr^{-1}}$.
According to the stellar initial mass fuction (IMF) in the solar neighbourhood
\citep{2002ASPC..285...86K},  
${\rm n_{ccsn}=5.3\times10^{-3}}$ core-collapse supernovae (CCSNe) occur
per unit mass of star formation\footnote{The number of stars in unit mass is derived by integrating the IMF, 
namely ${\rm \int^{m_u}_{m_l} \frac{d\,N}{d\,M} d\,M}$, 
where ${\rm \frac{d\,N}{d\,M}}$ corresponds to the IMF and ${\rm m_u}$ and ${\rm m_l}$
are the upper and lower masses.
We assume that ${\rm m_u=25~M_{\odot}}$ and ${\rm m_l=10~M_{\odot}}$ for
CCSNe (see Sec.~\ref{sec:model-2}).
The number of events in unit mass is summarized in Table~\ref{tab:nombre}.
For type-Ia supernova and neutron star merger, ${\rm n_{SNIa}}$ and ${\rm n_{NSM}}$ in the table are derived by 
integrating the delay time distributions from 0 to 14~Gyr.
In this article, the number of events in unit mass is denoted by 
${\rm n_{event}}$ (a lower-case letter), while ${\rm N_{event}}$ 
(an upper-case letter) represents the number of events in a time-step.
}.
The number of events in a time-step ${\rm \Delta t}$ at time $t$ 
($N_{\rm event}(t)$) is derived by
\begin{equation}
\label{eq:nombre}
N_{\rm event}(t) = \Psi(t) \, {\rm \Delta t} \, {\rm n_{event}}. 
\end{equation}
Thus, in average $N_{\rm ccsn} = \Psi \, {\rm \Delta t} \, {\rm n_{ccsn}} = 10^5 \, 10^{-3} \, 5.3\times10^{-3} \sim 0.53$ 
CCSNe occur in a time-step.
$N_{\rm ccsn}<1$ seems to represent no event, but there is a physical
meaning.
For instance, we can interpret 0.1 CCSNe as one CCSN occurs in the subsequent 
ten time-steps \citep[also see Section~3 of][]{2020MNRAS.493.3464P}.
In this study, we refer to an event as 'rare event' when 
the number of the source of elements in a time-step is less than unity.
Other sources than CCSNe are even more rare.
The occurrence of candidates of the r-process sites is smaller than that of CCSNe (see Section~\ref{sec:model-2-1} 
for assumptions about the r-process sites in this study).
For instance, for NSM, when the rate of NSM in the Milky Way 
\citep[${\rm 42^{+30}_{-14}~Myr^{-1}}$,][]{2019ApJ...870...71P}
and the delay time distribution \citep[DTD,][]{2019MNRAS.487.4847B}
are adopted, the number of NSM in unit mass is estimated to be 
${\rm n_{NSM} = 1.2\times10^{-5}}$.
Regarding collapsars, the number in unit mass
can be as small as ${\rm n_{col} = 5.3\times10^{-5}}$ 
if the occurrence is assumed to be 1~\% of that of CCSNe.

\begin{figure}
\begin{minipage}{0.45\linewidth}
\centering
\includegraphics[scale=0.41]{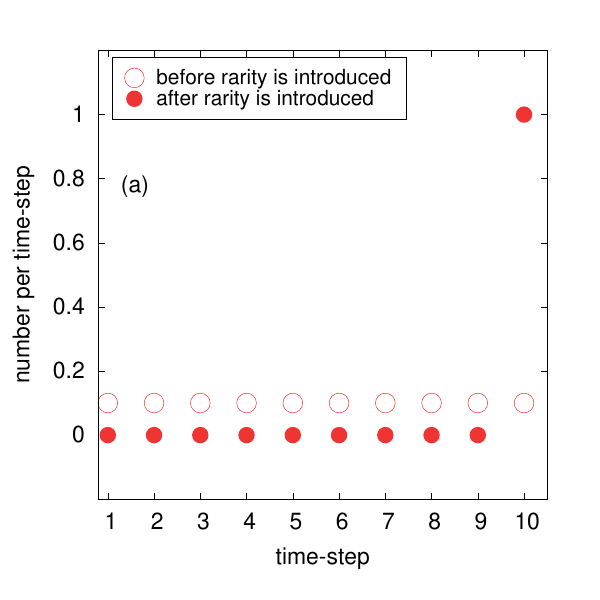}
\end{minipage}
\begin{minipage}{0.45\linewidth}
\centering
\includegraphics[scale=0.41]{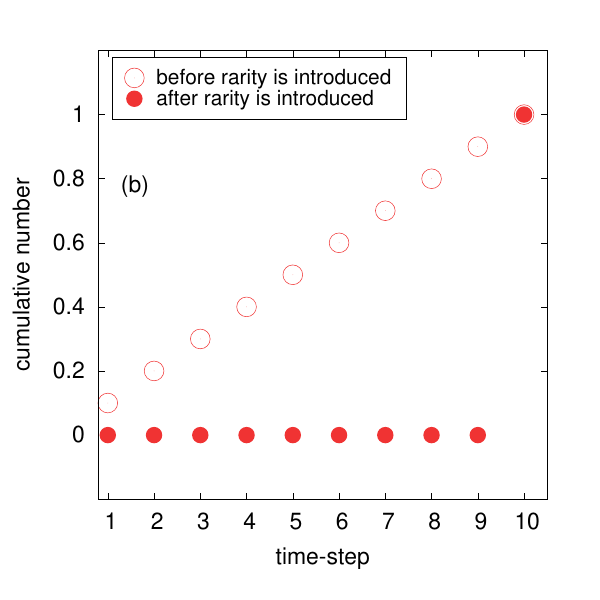}
\end{minipage}
\caption{The number of events in a sequence of ten time-steps (see text) is shown as
an example. (a) The number of events in each time-step. (b) The cumulative number.
In each panel, red open and filled circles represent the number of events before
and after the rarity is introduced, respectively.
}
\label{fig:app-rare}
\end{figure}

To introduce the concept of rarity into calculations,
we consider a sequence of ten time-steps, 
${\rm \Delta t_1, \Delta t_2, \cdots, \Delta t_{10}}$, as an example
(Figure~\ref{fig:app-rare}).
The number of events in each time-step ($N_{\rm event}$) is $0.1$.
In ${\rm \Delta t_1}$, $N_{\rm event}=0.1$.
Since the number of events is less than unity, the concept of rarity is introduced.
We put zero event in ${\rm \Delta t_1}$ and keep the 0.1 events.
In the next time-step (${\rm \Delta t_2}$), $N_{\rm event}=0.1$, so we again put
zero event in this time-step and keep the 0.1~events.
Since we have stored 0.1~events at ${\rm \Delta t_1}$,
there are $0.1+0.1=0.2$~events in the storage at ${\rm \Delta t_2}$.
We do the same operation for ${\rm \Delta t_3, \cdots, \Delta t_9}$.
When we get to ${\rm \Delta t_{10}}$, there are 0.9~events in the storage.
Since $N_{\rm event}=0.1$ at ${\rm \Delta t_{10}}$, we have $0.9+0.1=1$ event
in the storage and put the one event in ${\rm \Delta t_{10}}$.
The variation of $N_{\rm event}$ of this sequence and the cumulative number of
the events are shown in Fig.~\ref{fig:app-rare}.
The total numbers of the events at ${\rm \Delta t_{10}}$ before and after the rarity is introduced
are consistent.

We note that the rarity depends on both 
the mass of a system and the time-step.
When we set the time-step to be ${\rm \Delta t = 10^{-2}}$~Gyr,
about five CCSNe occur in the dwarf galaxy of ${\rm M_*=10^6~M_{\odot}}$
in a time-step, and thus CCSNe are not considered to be rare,
while the r-process events are rare.
In Appendix~\ref{sec:app-tstep}, we estimate 
the number of events in dSphs by
adopting ${\rm \Delta t = 10^{-1}}$ and 
${\rm 10^{-2}}$~Gyr .
In this study, the time-step is set to be ${\rm \Delta t = 10^{-3}}$~Gyr 
in order to take 
into account the ejecta from
massive stars and the precision of calculations.
Once the size of time-step is fixed, the number of events depends on 
the star formation rate.
The star formation history is reflected in the stellar mass of the system,
and thus
astrophysical events can be more rare in lower-mass galaxies.

\section{The application to dwarf galaxies}
\label{sec:3}
We apply our concept of rare events to two Local Group dSphs
(Sculptor and Sextans dSphs). 
These galaxies are similar in that stars have been formed most actively 
at the early time of the evolution of the galaxies and then 
the star formation stopped. 
Sculptor dSph is more massive than Sextans dSph. 
In this section, we firstly summarize the properties of the two dSphs
(Section~\ref{sec:3-1}). 
Then, we estimate the occurrence of each astrophysical source in the dwarf galaxies based on 
the star formation history (Section~\ref{sec:3-2}). 
In Section~\ref{sec:3-3}, the model for dwarf galaxies is presented.

\subsection{Summary of Sculptor and Sextans dSphs}
\label{sec:3-1}
\subsubsection{Sculptor dSph}
Sculptor dSph (mass in stars ${\rm M_*=3.9\times10^6~{\rm M}_{\odot}}$)\footref{foot:mass}
is the first dSph to be discovered in
1938 by Harlow Shapley \citep{1938BHarO.908....1S}.  
This dSph appears at a high latitude and observations are less affected
by the extinction. Thus, Sculptor dSph is one of the most studied
dwarf galaxies in the Local Group.

At first, this galaxy is recognized as a system similar to globular 
clusters \citep{1939PASP...51...40B}   
with no young star or gas.
Later, it is found that Sculptor dSph has an extended star formation
history compared to globular clusters and that the intermediate-age
population exists as well as the old ($>$10~Gyr) population
\citep[e.g.][and references therein]{2001MNRAS.327..918T}.
According to the star formation history derived by the colour-magnitude diagram,
the star formation rate peaked at the early time, declines with time 
and then turned down to almost zero around 8~Gyr ago 
\citep{2012A&A...539A.103D,2018MNRAS.480.1587S}.

In terms of the trends of the stellar abundance ratios,
including the [${\rm \alpha}$/Fe] ratios similar to those of the Galactic
halo stars at [Fe/H]~$\lesssim -1.8$, there is not a clear evidence for
the variation in the IMF \citep[e.g.][]{2019A&A...626A..15H}.
Comparing the abundance ratios of metal-poor stars in Sculptor dSph
and those of the Galactic stars, \citet{2015A&A...583A..67J}  
show that
most of the metal-poor stars (including extremely metal-poor stars) in 
Sculptor dSph follow trends of the ratios of the Galactic halo stars
and suggest a similar condition of the star formation at the early stage
of the galaxies.

At low metallicities, there is the scatter in the abundance ratios for
some of elements, which is interpreted as the result of the low occurrence
of supernovae \citep{2015ApJ...802...93S}. 
For instance, the [Ba/Fe] ratio varies among stars more than 
the uncertainty of the measurement at [Fe/H]~$< -1.8$,
perhaps due to rare events of the r-process \citep{2020A&A...634A..84S}.

With regard to the source of elements, the [Ba,Y/Mg] ratios as well as
the [Y,La,Nd,Eu/Ba] ratios of stars in this dSph suggest the contribution of AGB stars to the
enrichment of elements heavier than the Fe-peak at [Fe/H]~$> -2$
\citep{2020A&A...634A..84S,2021ApJ...915L..30S}.
The change in the slope of the Y abundance at [Ba/H]~$\sim-2$ is interpreted
as due to the existence of different production sites for these elements at low
metallicities \citep{2021ApJ...915L..30S}. 

\subsubsection{Sextans dSph}
Sextans dSph (${\rm M_*=7.0\times10^5~{\rm M}_{\odot}}$)\footref{foot:mass}
was discovered 
on UK Schmidt Telescope sky survey plate \citep[][]{1990MNRAS.244P..16I}. 
This dSph is known to have an extended structure \citep[tidal radius ${\rm r_t=83.2\pm7.1'}$,][]{2016MNRAS.460...30R}.
Also, the mass-to-light ratio can be about 120 times
greater than the solar value \citep{1994MNRAS.269..957H},
which is interpreted as Sextans dSph is dominated by dark matter.

This dSph seems not to have experienced or undergo strong tidal destruction
\citep[e.g.][]{2016MNRAS.460...30R,2017MNRAS.467..208O,2018A&A...609A..53C}.
Subcomponents of different velocities 
and metallicities might suggest mergers and/or accretions in the past
\citep[e.g.][]{2018MNRAS.480..251C}.

The star formation in Sextans dSph reached the peak 
at the early stage of the evolution \citep[e.g.][]{2009ApJ...703..692L}
and then have ceased within $\sim 1.3$~Gyr after the Big Bang 
\citep{2018MNRAS.476...71B}.
This galaxy is dominated by old stellar populations, such as horizontal
branch stars \citep[e.g.][]{1991AJ....101..892M}. 
According to an estimate of luminosity generated by SNe, most of the gas
in the galaxy seems to have been expelled due to SNe before the reionization
terminated \citep{2018MNRAS.476...71B}.

With regard to the chemical enrichment, the [${\rm \alpha}$/Fe] ratio
lower than those of the Galactic halo stars at higher metallicity
([Fe/H]~$\gtrsim -1.5$) suggests the contribution of type-Ia supernovae 
(SNe\,Ia) to the enrichment 
and perhaps the formation of a small number of massive stars 
\citep{2001ApJ...548..592S}. 
The ${\rm \alpha}$-knee (the point where the [${\rm \alpha}$/Fe] ratio starts to decrease with metallicity) is supposed to be located at [Fe/H]~$\sim -2$ 
\citep{2020A&A...641A.127R,2020A&A...642A.176T}.
\citet{2020A&A...642A.176T}  
point out that the metallicity at the knee of Sextans
dSph is close to those of more massive dSphs and that the efficiency of
star formation might be similar at the early time of the evolution 
among the dSphs.
Also, Sextans dSph might have another knee in [Mg/Fe] at 
[Fe/H]~${\rm \sim -2.5}$, 
which might reflect an accretion event or a merger in the past
\citep{2020A&A...641A.127R}.

At metallicities lower than the ${\rm \alpha}$-knee, the [${\rm \alpha}$/Fe] 
ratio of extremely metal-poor stars is less clear.
Analyses of high-resolution spectra of extremely metal-poor stars suggest the
[Mg/Fe] ratios lower than the average of the Galactic halo stars
\citep[e.g.][]{2009A&A...502..569A}. 
Stars of low [${\rm \alpha}$/Fe] ratios have been found in the Galactic
halo \citep[e.g.][]{2003ApJ...592..906I,2013ApJ...778...56C}. 
In the meanwhile, some studies suggest the [Mg/Fe] ratios similar to stars in the Milky Way
and an IMF similar to our Galaxy
\citep[e.g.][]{2020A&A...641A.127R,2022MNRAS.509.3626M}.
The measurement of Mg is likely to depend on 
absorption lines used in the analysis \citep{2020A&A...644A..75L}  
and the assumption about the local thermodynamic equilibrium 
\citep{2020A&A...641A.127R}.

As for elements heavier than the Fe-peak, 
the correlation between the abundance of Mg and Ba of stars in Sextans dSph
might suggest the r-process site without the large time delay relative to
CCSNe \citep{2020A&A...641A.127R,2022MNRAS.509.3626M}.

\subsection{The occurrence of astrophysical events in the dSphs}
\label{sec:3-2}
With the star formation history of individual galaxies derived by the 
colour-magnitude diagram, we estimate the occurrence of astrophysical sources 
in each time-step for Sculptor and Sextans dSphs as done in Sec.~\ref{sec:2}.
The assumptions about the chemical enrichment are described in Section~\ref{sec:model-2}.

\begin{figure*}
\centering
\includegraphics[scale=0.8]{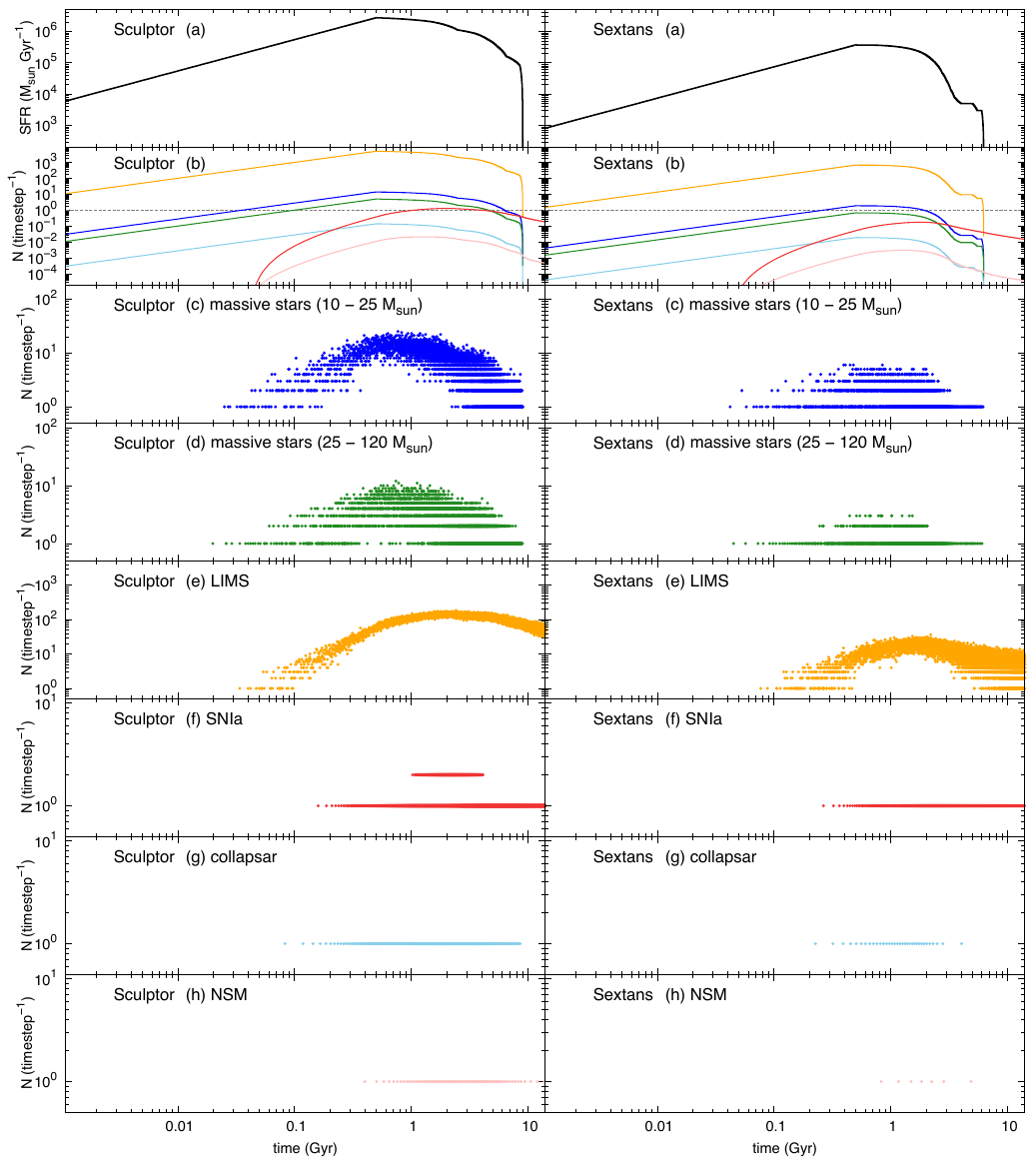}
\caption{The time variation of the star formation rate and the number of 
events in a time-step. 
Figures in left and right columns show the cases of Sculptor and Sextans dSphs,
respectively.
(a) The star formation histories of Sculptor and Sextans dSphs
taken from \citet{2012A&A...539A.103D}  
and \citet{2018MNRAS.476...71B}.  
(b) The number of astrophysical sources in a time-step if the rarity is not introduced. 
The colour of curves corresponds to the
astrophysical sources: massive stars of ${\rm 10 - 25~M_{\odot}}$ assumed to evolve into CCSNe (blue), massive stars of ${\rm 25 - 120~M_{\odot}}$ assumed to evolve into black holes (green), LIMS (orange), SNIa (red), collapsar (light blue) and NSM (pink). For the assumption about the evolution of massive stars, see Section~\ref{sec:model-2}. 
Horizontal dotted lines in figures (b) represent $N_{\rm event}(t) = 1$. Below this line, events are rare.
(c -- h) The number of massive stars (${\rm 10 - 25~M_{\odot}}$), massive stars (${\rm 25 - 120~M_{\odot}}$), LIMS, SNIa, collapsar and NSM that release the material in a time-step after the rarity is introduced. Each point shows the number of events in a given time-step.
}
\label{fig:nombre}
\end{figure*}

Figures~\ref{fig:nombre}a show the time variation
of star formation rate in each galaxy (\citealt{2012A&A...539A.103D}  
for Sculptor dSph and \citealt{2018MNRAS.476...71B}  
for Sextans dSph, respectively)\footnote{\citet{2012A&A...539A.103D}  
derive star formation rates for every age bin.
The star formation might not always be continuous between the adjacent age bins.
Generally, it is not easy to tell whether starbursts are continuous or not,
although the discontinuity of star formation may be reflected in the shape of metallicity distributions and abundance ratios.
In this study, we simply interpolate the star formation rates at 
a given stellar age.}.
Sculptor and Sextans dSphs have the peak of the
star formation at the early time of the evolution. 
After the peak, their star formation rates decrease and 
turn into almost zero several Gyrs ago.

Then, we estimate the number of events in a time-step at time $t$ 
($N_{\rm event}(t)$).
Blue, green and orange curves in Figs.~\ref{fig:nombre}b show 
the time variation of the number of massive stars and low- and intermediate-mass stars (LIMS) formed 
in a time-step.
The number of events in a time-step seems generally larger in Sculptor dSph 
that is more massive than Sextans dSph.
The total number of stars formed throughout the evolution of the dSphs
and that of SNIa and the r-process candidates are summarized in
Table~\ref{tab:nombre}.

\begin{table}
	\centering
        {\scriptsize
	\caption{The number of astrophysical sources in unit mass (${\rm n_{event}}$) and the total number of astrophysical sources throughout the evolution in Sculptor and Sextans dSphs. The galaxy mass in stars of the dSphs shown in the first row is taken from \citet[also see footnote~\ref{foot:mass}]{2013ApJ...779..102K}. For massive stars and LIMS, the number of stars formed in the dSphs is shown.}
	\label{tab:nombre}
	\begin{tabular}{llll} 
		\hline
		 & ${\rm n_{event}}$ & Sculptor & Sextans \\
		\hline
                galaxy mass in stars (${\rm M_{\odot}}$) & --- & $3.9\times10^6$ & $7.0\times10^5$ \\
		massive star (${\rm 10-25~M_{\odot}}$)& $5.3\times10^{-3}$ & $38571$ & $3446$  \\
		massive star (${\rm 25-120~M_{\odot}}$) & $1.9\times10^{-3}$ & $13827$ & $1235$  \\
		LIMS & $1.9$ & $13827579$ & $1235474$  \\
		SNIa & $1.3\times10^{-3}$ & $8641$ & $790$  \\
		NSM & $1.2\times10^{-5}$ & $86$ & $7$  \\
		collapsar & $5.3\times10^{-5}$ & $385$ & $34$  \\
		\hline
	\end{tabular}
}
\end{table}

As for SNIa and NSM, the explosive events occur certain time after 
the formation of white dwarfs and neutron stars, respectively.
We derive the number of SNIa and NSM that release heavy elements in a time-step
at time $t$ based on the DTDs and the star formation rate; 
\begin{eqnarray}
\label{eq:2}
N_{\rm event}(t) = R(t) \, {\rm \Delta t},
\end{eqnarray}
where $R(t)$ represents the rate of SNIa or NSM at time $t$ given by
\begin{eqnarray}
\label{eq:3}
R(t) = \int^t_0 \, \Psi(t')\,DTD(t - t') \,{\rm d}t'. 
\end{eqnarray}
$t'$ denotes the time of the star formation.
For SNe\,Ia, we assume $DTD \propto t^{-1}$ 
based on the observationally recovered delay-time distribution \citep[][and references therein]{2017ApJ...848...25M}.  
It is after the formation of the first white dwarfs ($\sim$0.04~Gyr) 
that SNe\,Ia start to appear.
At the early stage of the evolution of the galaxy (up to $\sim$1.0~Gyr),
the formula of the single degenerate scenario by \citet{2005A&A...441.1055G}  
is adopted.
With regard to NSM, we adopt the delay-time distribution proposed by
\citet{2019MNRAS.487.4847B}.  
Red and yellow curves in Figs.~\ref{fig:nombre}b show the time variation of 
the number of SNIa and NSM that occur in a time-step.

Regarding collapsars, we assume that the occurrence is 1~\% of that of CCSNe
(Section~\ref{sec:model-2-1}).
The time variation of the number of collapsars that occur in a time-step is 
shown in light blue curves in Figs.~\ref{fig:nombre}b.

Figs.~\ref{fig:nombre}b
show that the number of each source in a time-step can be less than unity
during the evolution of the dSphs.
In particular, the number of candidates of the r-process sites 
(NSM and collapsar)
is smaller than unity at any time.
When the number of each source is less than unity, we consider the source
to be rare and adopt the concept of the rarity as defined in Section~\ref{sec:2}.

Figs.~\ref{fig:nombre}c--h show the time variation of the number of each source that releases the material
in a time-step after the rarity is introduced.
In the model, each event releases the material and the chemical evolution proceeds
every time-step.
The number of collapsar in a time-step 
(Figs.~\ref{fig:nombre}g)
and that of NSM (Figs.~\ref{fig:nombre}h) are always one or zero. 

\subsection{Chemical evolution model for dwarf galaxies}
\label{sec:3-3}
\subsubsection{Global evolution}
We assume that a dSph evolves through the accretion of gas, the star
formation and the outflow.
There is no gas, star or heavy element in the galaxy 
at the beginning ($t=0$~Gyr).
The metal-free gas composed of H and He accretes on to the galaxy
at the infall rate $F(t) \propto {\rm e}^{-{\rm \alpha} t}$, 
where ${\rm \alpha~(Gyr^{-1})}$ is a constant referred to as the
time-scale of the gas accretion. 
Generally, dSphs do not have ongoing star formation, and 
there is not clear evidence for the presence of interstellar gas to 
form stars.
It is difficult to fully infer the time variation of the gas accretion in dSphs,
so we roughly assume that the accretion of gas on to the galaxy stops around the
time when the star formation rate turns into zero.
The gas in the galaxy is assumed to be instantaneously mixed.

Part of the gas that has accreted on to the galaxy turns into stars.
To derive the quantity of gas consumed by the star formation,
we refer to the number and mass of stars formed in a time-step after the rarity
is introduced (Figures~\ref{fig:global}a).
We discuss the time variation in Section~\ref{sec:4-1}.
As for the interplay between the accretion of gas and the star formation,
in our model the star formation rate is independent 
of the infall rate and the gas mass of the galaxy.
Since abundance ratios depend on the star formation
history, we think that adopting the star formation history derived from the
colour-magnitude diagram suits the purpose of this study.

With regard to the outflow, we assume that part of the interstellar gas is expelled 
by CCSNe\footnote{While the typical explosion energy of a single SNIa can be
comparable to that of a CCSN (about $10^{51}$~ergs),
massive stars can be formed in a group, such as star clusters and 
OB association. 
The stellar winds and CCSNe give heat and energy to the interstellar medium,
bubbles can be formed and the interstellar gas can be removed.
Following this scenario, we assume that the outflow occurs due to
CCSNe. We also note that the total number of SNe\,Ia is generally smaller than
that of CCSNe.}
. 
The outflow rate is determined as the amount of gas expelled from the galaxy due to the outflow in a time-step is proportional to the number of CCSNe that 
occur in the time-step.
The chemical composition of the gas expelled from the galaxy is
assumed to be consistent with that of the interstellar gas.
We note that dSphs can lose the gas through processes other than CCSNe\footnote{The outflow induced by the active galactic nuclei (AGN) may also affect the evolution of galaxies, however, the fraction of such dwarf galaxies is estimated to be smaller than about 2~\% \citep[e.g.][]{2021ApJ...922L..40L}. 
We focused on the case where the evolution of dSphs is not significantly affected by AGN.}, such as tidal interactions.
In the meanwhile, mechanisms that remove the gas from galaxies are not fully identified for individual dSphs.
Thus, we simply assume that the accreted gas is used for the star formation and expelled due to the outflow,
and that the mass of the gas in the galaxy turns into almost zero.

Since there is almost no gas to form stars in dSphs at the present universe,
the infall rate, the star formation rate and the outflow rate are normalised
as the total mass of gas that accretes on to the galaxy (${\rm M_{acc}}$)
is equal to the sum of the mass of gas that is turned into stars throughout
the evolution of the galaxy (${\rm M_{sfr}}$) and that expelled from the
galaxy because of the outflow (${\rm M_{out}}$) ; ${\rm M_{acc} \sim M_{sfr} + M_{out}}$.
Due to mass ejection from stars, the amount of gas in the galaxy at the
present universe predicted by the model does not completely become zero, but
it gets sufficiently low at the end of the evolution of the dSphs 
(see Figs.~\ref{fig:global}c).
The outflow rate at time $t$ is described as $o(t) = {\rm k} N_{\rm ccsn}(t) \frac{1}{\rm \Delta t} \frac{1}{\rm M_{acc}}$, where ${\rm k}$ is a constant referred to as the efficiency of the outflow.

The time-scale of the gas accretion (${\rm \alpha~Gyr^{-1}}$) and the efficiency of the outflow (${\rm k}$) are determined based on metallicity distributions
(see Figure~\ref{fig:global2}).
When a galaxy is assumed to lose the gas that includes heavy elements
through supernovae, the average metallicity of stars in the galaxy
predicted by the model gets low \citep[e.g.][]{1976ApJ...209..418H}.
Also, it is widely known that the accretion of gas is one of physical
processes that alleviate the overestimate of stars at the low-metallicity
tail of metallicity distributions \citep[e.g.][]{1975VA.....19..299L}.
We observe similar trends in metallicity distributions predicted by the model 
that includes rare events. 
The time-scale of the gas accretion and 
the efficiency of the outflow are roughly determined by comparing the shape
of the observed distributions with that predicted by the model.
Table~\ref{tab:inout} summarizes the determined ${\rm \alpha}$ and ${\rm k}$ 
and the representative values.
The value of ${\rm k}$ for Sextans dSph is larger than that for Sculptor dSph, 
because the average metallicity of Sextans dSph is lower than that of 
Sculptor dSph.

A dwarf galaxy acquires the gas through the gas accretion
and the mass ejection by stars, while part of the interstellar gas
is used for the star formation and expelled due to the outflow.
Thus, the evolution of gas in the dSph is described as follows;
\begin{eqnarray}
\label{eq:4}
\frac{{\rm d}\,m_{\rm g}}{{\rm d}\,t} &=& -\psi(t) + e_{\rm g}(t) + f(t) - o(t),
\end{eqnarray}
where $m_{\rm g}$ represents the mass of gas in the galaxy divided by the 
total mass of gas that has accreted on to the galaxy (${\rm M_{acc}}$).
All the factors are normalised by ${\rm M_{acc}}$ in calculations.
$e_{\rm g}$ is the rate of mass ejection by dying stars.

The evolution of mass fraction of element $i$ ($X_i$) is given by 
\begin{eqnarray}
\label{eq:5}
\frac{{\rm d}\,(m_{\rm g} X_i)}{{\rm d}\,t} &=& -\psi(t)\,X_i(t) + e_i(t) + f(t)\,{\rm X}_{i,{\rm in}} - o(t)\,X_i(t),
\end{eqnarray}
where ${\rm X}_{i,{\rm in}}$ is the mass fraction of element $i$ in the accreting gas.
$e_i(t)$ represents the rate of ejection of element $i$ by stars and events.

\begin{table}
	\centering
	\caption{The representative quantities of the rates of the gas accretion and the outflow in the model. ${\rm \alpha}$ is a constant that determines the time-scale of the accretion of gas. ${\rm t_{in}}$ represents the time when the accretion of gas stops. ${\rm k}$ is the efficiency of the outflow that is assumed to be a constant. 
We vary ${\rm k}$ by $1.0\times10^3$ for Sculptor dSph
and $2.0\times10^3$ for Sextans dSph, respectively. ${\rm \alpha}$
is determined by varying the value by $0.1$.
The representative values are chosen based on comparisons between the observed and predicted metallicity distributions.}
        {\scriptsize
	\begin{tabular}{lll} 
		\hline
		 & Sculptor & Sextans \\
		\hline
		${\rm k}$, ${\rm \alpha~(Gyr^{-1})}$ & ${\rm k=5.0\times10^3, \alpha=0.9-1.1}$ & ${\rm k=8.0\times10^3, \alpha=1.3-1.7}$ \\
		                                    & ${\rm k=6.0\times10^3, \alpha=0.8-1.0}$ & ${\rm k=1.0\times10^4, \alpha=1.2-1.4}$ \\
		representative ${\rm k}$, ${\rm \alpha~(Gyr^{-1})}$ & ${\rm k=5.0\times10^3, \alpha=1.0}$ & ${\rm k=8.0\times10^3, \alpha=1.4}$ \\
		${\rm t_{in}~(Gyr)}$ & $8.5$ & $6.0$ \\
		\hline
	\end{tabular}
	\label{tab:inout}
}
\end{table}

\subsubsection{Chemical enrichment}
\label{sec:model-2}
The sources of the chemical enrichment in the model are massive stars, LIMS, 
SNe\,Ia and the r-process candidates.
The number of events that release the ejecta at a given time follows that
estimated in Sec.~\ref{sec:3-2}.
Although the possibility of the variation in the IMF among different systems
is not rejected \citep[e.g.][for a review]{2013pss5.book..115K},
the [${\rm \alpha}$/Fe] ratio of stars
in Sculptor seems generally similar
to that of the Galactic halo stars, 
and the clear evidence for the variation in the IMF during the evolution
of the dwarf galaxies is not found from the viewpoint of the 
[${\rm \alpha}$/Fe] ratio. 
Thus, we adopt the IMF in the solar neighbourhood to the dSphs.
For each star, the mass is randomly given by weighting with the IMF 
\citep{2002ASPC..285...86K}  
in the range of ${\rm 0.08 - 120~M_{\odot}}$.
Once the mass is given, the age of the star $\tau_M$ is also determined based on stellar models by \citet{1992A&AS...96..269S}
\footnote{We note that a rotating star may have a longer lifetime than a non-rotating
star of the same mass, since rotational mixing supplies the source
to nuclear burning regions. Ideally, this effect of rotation on 
stellar age should be taken into account in calculations. 
At the same time, 
other ingredients in the model, such as yields and star formation rates, 
also include uncertainties.
In this study we adopt Schaller et al. data to the models 
rather than taking into account the effect of rotation.
}.
A star that is formed at time $t_0$ releases heavy elements at $t=t_0+\tau_M$.
The metallicity of a star is determined based on that of the interstellar 
gas. 

Regarding massive stars ($M > 10~{\rm M}_{\odot}$), total yields provided by 
\citet{2018ApJS..237...13L}  
are adopted.
Mass, metallicity and initial rotating velocity range 
$13~{\rm M}_{\odot} < M < 120~{\rm M}_{\odot}$, 
[Fe/H]$=0, -1, -2$ and $-3$
and $v_{\rm ini}=0, 150$ and $300~{\rm km\,s^{-1}}$, respectively.
The initial rotating velocity is randomly given 
by weighting with the initial distribution of rotating velocities\footnote{The fractional contribution of 
yields for rotating massive stars is considered to be the number fraction of 
stars of different rotating velocities. 
In \citet{2018MNRAS.476.3432P}, the fractional contribution of the yields for
rotating massive stars is determined based on the trends of the ${\rm {}^{14}N}$
abundance and the s-nuclei. The abundance ratio predicted by chemical evolution
models depends on the formalism of the model and the assumptions. In this study,
we add a small modification to the fractional contribution proposed by \citet{2018MNRAS.476.3432P}
through comparisons between the trends in abundance ratios of the Milky Way stars
and those derived by the infall model described in Appendix~\ref{app:2}
and then adopt the fractional contribution to the model for dwarf galaxies by assuming
that it is not varied among different systems. The modification is not drastic and 
it does not affect the global results.}
in \citet{2018MNRAS.476.3432P} 
in the range of ${\rm 0 - 300~km\,s^{-1}}$ on the assumption that 
the initial distribution of rotating velocities does not vary among different
systems.
Thus, massive stars of an identical metallicity can have different initial
rotating velocities.
Stellar winds during the evolution are included in the total yields.
For the explosion of the progenitors, the mixing and fall back is adopted and 
a single star is assumed to eject ${\rm 0.07~M_{\odot}}$ of ${\rm {}^{56}Ni}$.
Stars of $M > 25~{\rm M}_{\odot}$ are assumed to collapse into black holes
and contribute to the chemical enrichment only through the stellar winds.

LIMS ($M < 10~{\rm M}_{\odot}$) are assumed to evolve into AGB stars.
The total yields are obtained from the FRUITY database \citep{2015ApJS..219...40C}.
Mass and metallicity range $1.3~{\rm M_{\odot}} \leq M \leq 6~{\rm M_{\odot}}$ and 
$-2.15 \leq$ [Fe/H] $\leq 0.15$, respectively.
For LIMS we assume that non-rotating stars are dominant.

With regard to stars of the mass range 
$8~{\rm M_{\odot}} \lesssim M \lesssim 10~{\rm M_{\odot}}$,
massive AGB stars can contribute to the enrichment of nuclides
lighter than the Fe-peak (e.g. ${\rm {}^{14}N}$) through hot bottom burning.
Also, trans-iron elements can be produced through thermal pulses. 
However, the fate of stars in the mass range and the yields of massive
AGB stars depend on the assumptions in the models, 
such as mass loss and convection, and nuclear reaction rates
\citep[][for a review]{2017PASA...34...56D}.
In addition, a small fraction of stars in this mass range 
(about $2-5$~\% of CCSNe, \citealp{2011ApJ...726L..15W},  
\citealp{2015MNRAS.446.2599D}) 
can reach the explosion known as electron capture supernova 
\citep[ECSN,][]{1980PASJ...32..303M,1984ApJ...277..791N},
which can produce light trans-iron elements.
Although a supernova whose characteristics agree with those of ECSN
has been found \citep{2021NatAs...5..903H}  
and that the nucleosynthesis has been investigated \citep[e.g.][]{2011ApJ...726L..15W},
the mass range of the progenitors
is supposed to be narrow (0.1 -- 0.2~${\rm M_{\odot}}$, \citealp{2015MNRAS.446.2599D}, 
\citealp{2019ApJ...885...33H}) 
and depend on metallicity.
For these reasons, in this study, yields for massive AGB stars 
and ECSN are not included. 

For stars of ${\rm M}_{\rm L,up} < M < 10~{\rm M}_{\odot}$ where ${\rm M}_{\rm L,up}$ is 
the upper mass of the LIMS yields, the total yields are obtained by
\begin{equation}
y_i(M) = \frac{y_i({\rm M}_{\rm L,up})}{E({\rm M}_{\rm L,up})} E(M),
\end{equation}
where $E(M)=M-M_{\rm rem}$ and $M_{\rm rem}$ is the remnant mass
derived by the interpolation between the remnant mass of star of
6~${\rm M_{\odot}}$ and that of 10~${\rm M_{\odot}}$.
To obtain total yields of stars of ${\rm 10~M_{\odot}} < M < 13~{\rm M_{\odot}}$, 
we interpolate the yields of 10~${\rm M}_{\odot}$ and those of 13~${\rm M}_{\odot}$.
Total yields of stars of $M < 1.3~{\rm M_{\odot}}$ are derived by the extrapolation.

SN\,Ia is a major source of the iron-peak elements.
While comparisons between abundance ratios of stars and those
predicted by chemical evolution models suggest that 
SNe\,Ia of different classes might contribute to the chemical enrichment
in the solar neighbourhood and dSphs
\citep[e.g.][]{2019ApJ...881...45K,2020ApJ...895..138K,2021MNRAS.503.3216P},
the explosion mechanism is still under debate.
We adopt the yield provided by \citet{1999ApJS..125..439I}.

\subsubsection{Chemical enrichment via the r-process}
\label{sec:model-2-1}
The candidates and the contribution of the r-process sites to the chemical
enrichment are actively discussed \citep[e.g.][for a review]{2021RvMP...93a5002C}. 
Roughly speaking, collapses of massive stars and compact 
binary mergers have been proposed to be the dominant sites.

With regard to the latter, the production of the r-process material in
neutron star -- neutron star \citep[][]{1982ApL....22..143S} 
and neutron star -- black hole collisions \citep[][]{1974ApJ...192L.145L}
has been theoretically suggested. 
The nucleosynthesis through the r-process in NSM is supported by
the detection of the gravitational wave signal
\citep[GW170817,][]{2017PhRvL.119p1101A}  
and the associated electromagnetic 
sources \citep[AT 2017gfo, e.g.][]{2017Natur.551...67P}.
While the ejecta mass is supposed to depend on the property of the
progenitors \citep[e.g.][for the case of dynamical ejecta]{2013ApJ...773...78B}, 
the mass of ejecta estimated for GW170817/AT 2017gfo
\citep[${\rm \gtrsim 0.03~M_{\odot}}$, e.g.][]{2017ApJ...850L..37P,2017Natur.551...67P,2017PASJ...69..102T}
suggests that NSM might be a dominant source of the r-process elements in the
solar system \citep[e.g.][]{2017Natur.551...80K}. 
Also, trends of Mg and Ba abundances of stars in dwarf galaxies 
suggest that the production of Ba might be delayed relative to Mg
at [Fe/H]~$< -1.6$,
which has been interpreted as NSMs contribute to the enrichment
through the r-process 
\citep[e.g.][]{2018ApJ...869...50D}.  

As for the collapse of massive stars, since the r-process requires a
large flux of neutrons, supernova explosions have been supposed to be 
the candidate.
CCSNe might produce light trans-iron elements,
however, they might not be a main production site of heavier elements 
if the entropy is lower than required \citep[e.g.][]{2013ApJ...770L..22W}.

In the meanwhile, 
physical conditions that meet the nucleosynthesis through the r-process
are discussed and supernovae explosions that evolve from cores with
rapid rotation and strong magnetic field are suggested as an r-process site
\citep[e.g.][]{1966ApJ...143..626C,1972ApJ...173..195S,1976ApJ...204..869M,1985ApJ...291L..11S,2006ApJ...642..410N,2012ApJ...750L..22W}. 
Also, spinning black holes with jets, called as collapsars, 
have been studied as failed supernovae 
\citep[e.g.][]{1983ApJ...269..281B} 
and then proposed 
and investigated as candidates for progenitors of long gamma-ray bursts
\citep[][]{1993ApJ...405..273W,1995AdSpR..15e.143H,1996A&A...305..839J,1999ApJ...524..262M},
which may be associated with luminous supernovae
\citep[e.g.][]{1998Natur.395..670G}. 
Collapsars are supposed to produce a wide range of the trans-iron elements 
\citep[e.g.][]{2019Natur.569..241S}. 
While the frequency of the event is low (about 0.1 -- 1\% of the occurrence 
of CCSNe), collapsar may be able to explain about 80~\% of the solar Eu
abundance.

In terms of stellar abundances, the [Eu/Mg] ratio of stars in the Milky Way 
and Sculptor 
dSph is distributed around zero, which can be interpreted as Eu 
(a representative element that is mainly produced by the
r-process) is produced at an almost same 
time-scale as that of CCSNe \citep[e.g.][]{2019A&A...631A.171S}.

It is under debate whether the r-process proceeds at multiple sites.
From the viewpoint of the chemical evolution,
a decreasing trend of the [Eu/Fe] ratio
of the Galactic stars at [Fe/H] $> -1$ seems not to be fully
explained by chemical evolution models if NSM alone is assumed
to be the dominant r-process site
\citep[e.g.][]{2018IJMPD..2742005H,2019ApJ...875..106C}. 
With regard to low metallicities, 
the delay time of NSM as short as $\lesssim 10$~Myr is preferred to 
explain the trend of the [Eu/Fe] ratio of stars in the Milky Way
with theoretical models
\citep[e.g.][]{2004A&A...416..997A,2004NewAR..48..861D,2014MNRAS.438.2177M},
while the dispersion in the [Eu/Fe] ratio of the Galactic stars seems to be explained
with NSM when the chemical evolution of the Milky Way is 
modelled in a cosmological context where galaxies are formed through
mergers of sub-haloes of different masses and star formation efficiencies 
\citep{2018ApJ...865...87O,2021MNRAS.505.5862W}.
In terms of the chemical abundance of stars, the abundance pattern
of r-rich metal-poor stars is almost consistent with the solar r-process
pattern for elements of atomic number ${\rm Z \ge 63}$,
while for lighter trans-iron elements there is the variety
\citep{2008ARA&A..46..241S},  
which suggests that
light trans-iron elements are produced at multiple sites.

Since the contribution of each candidate of the r-process is still
open to discussion and that both compact binary mergers and 
rare collapses of rotating massive stars can be the r-process sites,
in this study 
we consider NSM and collapsar to be the main sites of the r-process. 

The occurrence of NSM is estimated based on the delay-time distribution
(equations~\ref{eq:2} and \ref{eq:3}).
With regard to collapsar, 
owing to keeping angular momentum due to less mass loss and 
the formation of larger helium cores, the occurrence is supposed to be more
frequent at low metallicities.
According to the rate of long gamma-ray bursts,
about 1~\% of type~Ibc supernovae can launch jets and produce long gamma-ray 
bursts at the local universe and the fraction can increase with 
redshift \citep[][]{2022ApJ...932...10G}. 
While the rate of long gamma-ray bursts can vary with metallicity and redshift,
the occurrence at low metallicities and the progenitors are still discussed 
\citep[e.g.][]{2019A&A...623A..26P,2021Galax...9...79R}.
In this study, we assume that the occurrence of collapsar is 1~\% of 
that of CCSNe \citep[][]{1999ApJ...524..262M}. 

Regarding the mass of each element ejected via a single r-process event, 
the yields for the candidates of the r-process sites depend on the assumptions 
in the nucleosynthesis calculations. 
The impact of nuclear data on the yields is also discussed 
\citep[e.g.][]{2023MNRAS.523.2551K}. 
With regard to NSM, for instance, the evolution of binary systems depends 
on mass of the merging compact objects 
\citep[e.g.][for a review]{2013CQGra..30l3001B} 
and also the mass of the ejecta is likely to depend on the mass ratio.
The yields for collapsars generally depend on the angular momentum 
and the viscosity of the accretion disc, and
it is still under debate whether heavier trans-iron elements are produced
by collapsars \citep[][]{2022arXiv221203958F}. 
In addition, analyses of light curves of long gamma-ray bursts suggest 
the production of a large amount of ${\rm {}^{56}Ni}$ through luminous and 
energetic supernovae (hypernovae),
but mechanisms and scenarios where both ${\rm {}^{56}Ni}$ and heavy trans-iron
elements are produced are not fully understood \citep[][]{2021RvMP...93a5002C}.

In this study, we simply assume that 
the abundance pattern of the yields of NSM and collapsar are consistent with
the solar r-process pattern. The yields for NSM and collapsar
(${\rm Y}_{i, \rm{NSM}}$ and ${\rm Y}_{i, \rm{col}}$, respectively)
are described as follows; 
\begin{eqnarray}
{\rm Y}_{i,{\rm NSM}} &=& (1-{\rm A_r})\cdot({\rm A_{NSM}}\,{\rm X}_{i,\odot}\,{\rm f}_{i,{\rm r}}), \nonumber \\ 
{\rm Y}_{i,{\rm col}} &=& {\rm A_r}\cdot({\rm A_{col}}\,{\rm X}_{i,\odot}\,{\rm f}_{i,{\rm r}}),
\label{eq:7}
\end{eqnarray}
where ${\rm A_{NSM}}$ and ${\rm A_{col}}$ are set to be $7.0\times10^4$ and
$1.5\times10^4$, respectively, as the solar Eu abundance
([Eu/H] $=0$ at [Fe/H] $=0$) is explained with the chemical evolution model
for the Galactic disc (see Appendix~\ref{app:2}) 
where either NSM or collapsar is
considered to be the r-process site. 
${\rm A_r}$ represents the contribution of collapsars 
to the solar Eu abundance.
For instance, when all of the solar Eu is assumed to be produced by collapsars 
(NSMs), ${\rm A_r=1\,(0)}$.
Whether the contribution of the r-process sites to the chemical enrichment
is universal among different systems or not is not clear.
Thus, we allow ${\rm A_r}$ to vary among dwarf galaxies and set to be
${\rm A_r=0.3}$ for each of Sculptor and Sextans dSphs.
The explanations about the contribution of the r-process sites 
in dSphs are found in Appendix~\ref{app:2}. 
${\rm f}_{i,{\rm r}}$ in equation \ref{eq:7} represents the r-component of the solar system.
We adopt the r-component derived with the chemical evolution model 
where the contribution of rotating massive stars through the weak s-process
is taken into account \citep{2020MNRAS.491.1832P}.
Global results of this study seem not to be changed when the 
r-fraction proposed by \citet{2008ARA&A..46..241S} 
is adopted.
Note, however, that the s- and r-components can be varied when
different methods are adopted. 
In addition, multiple production sites (e.g. ECSN) are proposed for 
light trans-iron elements, and thus comparisons between physical quantities
predicted by the model
with observational data have to be done carefully.

\section{Results}
\label{sec:4}
\subsection{Global properties of dwarf spheroidal galaxies}
\label{sec:4-1}
Using the chemical evolution model for dwarf galaxies, 
we predict global properties and chemical abundance for Sculptor and Sextans dSphs.
The representative time-scale of the gas accretion and the efficiency of the 
outflow in Table~\ref{tab:inout} are adopted.
We recall that it is average quantities that the model predicts.
While dSphs evolve through a number of physical processes, the model
includes only a small number of assumptions, and thus observational data
are not always fully reproduced by quantities derived with the model.
Nevertheless, the model is practical to see the impact of each source
of elements on the chemical abundance.

\begin{figure*}
\centering
\includegraphics[scale=1]{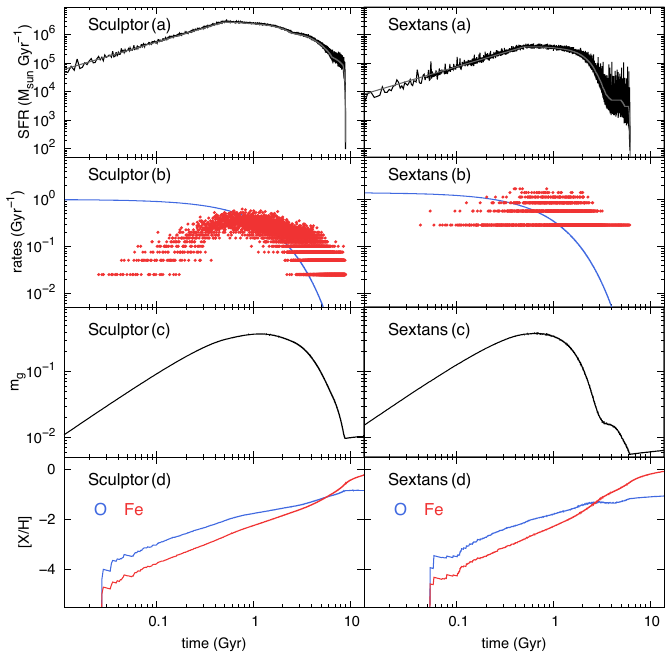}
\caption{The time variation of properties of Sculptor and Sextans dSphs 
(left and right panels, respectively) predicted by model A in Table~\ref{tab:model}. 
The panels show the time variation of 
(a) the modelled star formation rate after the rarity is introduced
(black curves),
(b) the rates of the gas accretion (blue curves) and the outflow (red dots),
(c) the mass of gas in the galaxies and 
(d) the abundances of oxygen (blue curves) and iron (red curves) 
in terms of [O/H] and [Fe/H], respectively.
In panels (a), the star formation history of Sculptor dSph \citep[][]{2012A&A...539A.103D} and that of Sextans dSph \citep[][]{2018MNRAS.476...71B} derived from the colour-magnitude diagram is also shown in grey curves.
For panels (b) and (c), the quantities are scaled to the total mass of the gas that
accretes on to the galaxy throughout the evolution.
}
\label{fig:global}
\end{figure*}

In Figs.~\ref{fig:global}a, the time variation of the star formation rate
in Sculptor and Sextans dSphs after the rarity is introduced is shown
(black curves).
Generally, the rates follow the star formation history derived from
the colour-magnitude diagram (grey curves).
When the rarity is introduced, the star formation rates oscillate. 
At the early and late time, the amplitude seems large, because the number of 
both massive stars and LIMS is small due to the low star formation rate
and that stars are formed sporadically.
When more stars are formed at a given time and/or 
the mass of the 
individual stars is larger, the star formation rate tends to be high.
Also, the amplitude of the oscillation in the star formation rate 
of Sextans dSph seems to be larger than that of Sculptor dSph.
The dependence of the fluctuation in the star formation rate on the
galaxy stellar mass has been reported for nearby star-forming galaxies
\citep[e.g.][]{2012ApJ...744...44W}.

Figs.~\ref{fig:global}b show the time variation of the rates of gas inflow 
and outflow scaled to the total mass of gas that accretes on to the galaxy 
throughout the evolution (${\rm M_{acc}}$). 
The gas continuously accretes on to the dSphs. 
Since the outflow occurs due to CCSNe, the outflow rate reflects 
the number of CCSNe (Figs.~\ref{fig:nombre}c). 
The outflow rate of Sextans dSph seems generally higher than that of Sculptor
dSph, mainly due to the high efficiency of the outflow.

Figs.~\ref{fig:global}c show the time variation of the mass of gas 
in the galaxy scaled to ${\rm M_{acc}}$. 
Since the gas accretes on to the galaxy and the gas is consumed by 
the star formation and the outflow, the mass of gas in the dSphs increases 
at the early time of the evolution and then decreases with time. 
The mass of gas in the galaxy does almost not change with time 
after the star formation stops in the galaxies,
while material is returned from LIMS.

In Figs.~\ref{fig:global}d, we show the evolution of the abundance of 
oxygen and iron in the dSphs as the representatives of metallicity. 
In general, the abundance increases with time. 
At the very early time of the evolution of the dSphs, 
the abundance oscillates due to the ejection from stars, the dilution of gas due to the accretion of the metal-free gas
and the outflow induced by CCSNe.
Oxygen is mainly produced by massive stars, and thus 
the [O/H] ratio generally increases during the period where stars are formed.
Iron is produced by SNe Ia as well as massive stars. 
Since progenitors of SNe Ia are white dwarfs and the explosive events 
occur certain time after the formation of white dwarfs, 
SNe Ia can enrich the interstellar medium after the star formation stops. 
Thus, the iron abundance continues to increase even at a later stage of 
the evolution of the dSphs.

\begin{figure}
\centering
\includegraphics[scale=1]{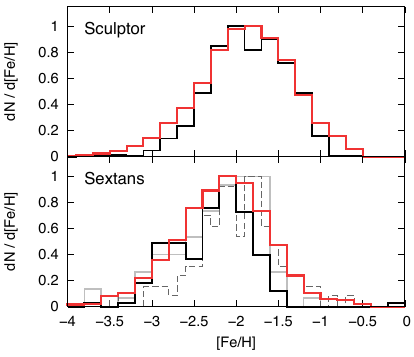}
\caption{Metallicity distributions of Sculptor dSph (upper panel) and 
Sextans dSph (lower panel). In the panels, black histograms show observed
distributions taken from \citet[][references therein]{2013MNRAS.434..471R} for Sculptor dSph 
and \citet[][]{2010A&A...513A..34S} for Sextans dSph.  
Grey histograms in the lower panel
are metallicity distributions of the central region of Sextans dSph
measured with medium- \citep[][dashed line]{2011ApJ...727...78K} and 
high-resolution spectra \citep[][solid line]{2020A&A...642A.176T}, 
respectively.
Red histograms are model predictions.
}
\label{fig:global2}
\end{figure}

In Fig.~\ref{fig:global2}, we compare metallicity distributions of the dSphs 
with those predicted by the model.
The observational data are referred from the literature as the difference
in the radial extent of observational data used to analyse the
metallicity distributions and the star formation history of each dSph
gets small as much as possible\footnote{Regarding the star formation history
and the metallicity distribution of Sextans dSph, \citet{2018MNRAS.476...71B}
focus on the star formation history of the inner region (central 
$34'\times27'$~region) of the dSph.
A metallicity distribution of the central region ($21'.4$) is also derived 
by \citet{2011ApJ...727...78K}. As the authors discuss in the paper,
stars in the central region tend to have high metallicities.
To constrain the infall rate in the model, we compare a metallicity
distribution that includes stars in the outer region presented by
\citet{2010A&A...513A..34S} with model predictions.
The average metallicity of the metallicity distribution of
\citet{2011ApJ...727...78K} is roughly consistent when the efficiency
of the outflow is assumed to be $k \sim 6\times10^3$ and we observe
similar trends in the predicted abundance ratios to those shown in
Sec.~\ref{sec:6}.
In the lower panel of Fig.~\ref{fig:global2}, we show metallicity
distributions of Sextans dSph by different previous works for reference.
For Sculptor dSph, when the model is compared to the metallicity
distribution that includes a larger number of stars 
\citep[][]{2023A&A...675A..49T}, the representative ${\rm k}$ and ${\rm \alpha}$
are ${\rm k = 5.0\times10^3}$ and ${\rm \alpha = 0.8~Gyr^{-1}}$ and
we observe the same trends as those in this article.}
.
We note that the shape of the predicted metallicity 
distributions is not completely consistent with that of the observed ones. 
As we stressed above, dwarf galaxies evolve through 
complex processes, such as mergers and stellar feedback. 
Regarding observational data, the shape of metallicity distributions 
of a galaxy can be different when the measurement method of metallicity 
and the observed area are changed \citep{2013MNRAS.434..471R,2020A&A...642A.176T}.
Considering the uncertainty, we think that the model captures
the global shape and the average metallicity of 
the observed metallicity distributions.

As explained above, after the star formation stops in the dSphs, 
low mass stars can return iron that they originally had
at the formation to the interstellar medium where only a small
amount of gas remains.
Also, SNe\,Ia release a large amount of iron.
Thus, the Fe abundance of the interstellar medium predicted by the model 
gets as high as [Fe/H]~$\sim 0$ (Figs.~\ref{fig:global}d).
On the other hand, the chemical abundance of old stars reflects the 
chemical composition of gas that formed the stars.
Since there is no current star formation in Sculptor and Sextans dSphs,
the modelled metallicity distribution of the dSphs extends up to 
low metallicities ([Fe/H]~$\sim -0.5$) relative to the Fe abundance of the 
interstellar medium.

\subsection{Abundance ratios}
\label{sec:6}
In this section, we compare abundance ratios of stars in Sculptor and Sextans
dSphs with those predicted by the model.
We again note that the model used in this study includes only a small number of 
assumptions and that it predicts average quantities. 
For instance, concerning the source of elements,  
the chemical abundance pattern of an ultra metal-poor star in Sculptor
dSph suggests the presence of SNe of explosion energies as high as
$10^{52}$~erg \citep[][]{2021ApJ...915L..30S},  
while the model used in the current study does not 
include yields for high-energy SNe derived from nucleosynthesis calculations.
If high-energy SNe frequently occur at low metallicities, 
observational data may not be always completely reproduced
by the model.

Nevertheless, the advantage of the model is that it is useful to observe 
the contribution of each source of elements to the chemical enrichment. 
We prepare models that include different sources of elements as summarized
in Table~\ref{tab:model}. 
Physical quantities other than the source of elements, such as the rates
of the accretion of gas and the outflow, the IMF and the initial distribution
of rotating velocities, are not varied among models A -- F.
In this article, we focus on a couple of representative elements.

\begin{table}
	\centering
	\caption{Summary of models.}
	\label{tab:model}
	\begin{tabular}{ll} 
		\hline
		 model & sources of elements \\
		\hline
		A & RMS + LIMS + SNIa + R (NSM + collapsar) \\
		B & no RMS + LIMS + SNIa + R (NSM + collapsar) \\
		C & RMS + LIMS + SNIa + R (NSM) \\
		D & RMS + LIMS + SNIa + R (collapsar) \\
		E & RMS + LIMS + SNIa + no R \\
		F & no RMS + LIMS + SNIa + no R \\
		\hline
	\end{tabular}
\end{table}

\subsubsection{The contribution of each astrophysical source to the chemical evolution}
\begin{figure}
\centering
\includegraphics[scale=1]{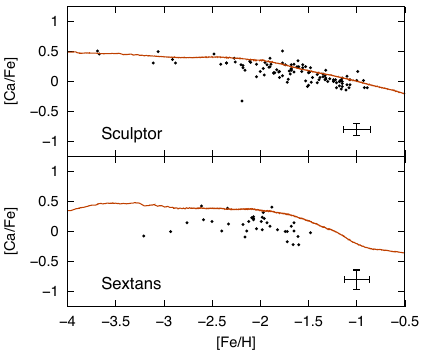}
\caption{The evolution of the [Ca/Fe] ratio in Sculptor and Sextans dSphs (upper and lower panels, respectively).
Black dots are the abundance ratio of stars in the dSphs. 
The observational data are taken from \citet{2017A&A...608A..89M}  
and \citet{2019A&A...626A..15H}  
for Sculptor dSph, and \citet{2020A&A...642A.176T}  
for Sextans dSph.
The typical uncertainty about the data is shown at the lower right of the panels.
Orange curves are the ratio predicted by model A for each dSph.
}
\label{fig:ratio1}
\end{figure}

The ${\rm \alpha}$-elements are mainly produced by massive stars, 
while iron is produced by SNe Ia as well as massive stars. 
Since SNe Ia occur after the formation of white dwarfs, 
the ratio of the abundance of the ${\rm \alpha}$-elements to that of iron 
([${\rm \alpha}$/Fe]) 
depends on the history of the star formation in a galaxy. 
In Figure~\ref{fig:ratio1}, we show the [Ca/Fe] ratio 
as a representative. 
The ratio of stars in Sculptor dSph shows a decreasing trend with 
increasing metallicity at higher metallicities ([Fe/H]~$\gtrsim -2$, \citealt{2019A&A...626A..15H}).
Generally speaking, at the early time of the evolution of a galaxy, 
massive stars produce both the ${\rm \alpha}$-elements and iron. 
Once SNe\,Ia start to contribute to the enrichment, 
the [${\rm \alpha}$/Fe] ratio declines with increasing metallicity
\citep[e.g.][]{1979ApJ...229.1046T}. 
This decline in the observed abundance ratios might partly be due to 
the small number of massive stars formed in the galaxy \citep{2001ApJ...548..592S}.  

Fig.~\ref{fig:ratio1} shows that model A for Sculptor dSph seems to 
roughly explain the trend of the [Ca/Fe] ratio of stars 
in Sculptor dSph. 
Stars in Sextans dSph show generally low [Ca/Fe] ratios compared to the
ratio predicted by the model, but considering the uncertainty about 
the observational data,
model A for Sextans dSph seems to roughly capture the trend of 
the average of the [Ca/Fe] ratio of the stars.

\begin{figure}
\centering
\includegraphics[scale=1]{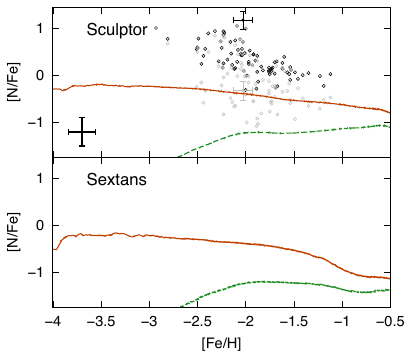}
\caption{The evolution of the [N/Fe] ratio of Sculptor dSph (upper panel) 
and Sextans dSph (lower panel) as a function of metallicity.
Black open circles in the upper panel are the abundance ratio of stars in 
Sculptor dSph taken from \citet{2016A&A...585A..70L}  
who have measured the N abundances
based on medium-resolution spectra. 
The typical uncertainty is shown at the lower left of the panel.
A black dot with error bars is the 
abundance ratio of a carbon-enhanced metal-poor star 
without the overabundance of heavy trans-iron elements (CEMP-no star) 
in the dSph measured with the high-resolution spectra \citep{2015A&A...574A.129S}. 
Grey open circles and dot are original [N/Fe] ratios of the stars analysed 
by \citet{2016A&A...585A..70L} and \citet{2015A&A...574A.129S}, respectively, estimated with stellar evolution models\protect\footnotemark.
Orange and green curves are the ratio predicted by models A and B, respectively.
}
\label{fig:ratio2}
\end{figure}

To investigate the impact of rotating massive stars on the
chemical abundance, we examine the evolution of the [N/Fe] ratio.
In Figure~\ref{fig:ratio2}, we observe a decreasing trend in the [N/Fe] ratio 
of red giant branch stars in Sculptor dSph (black open circles) 
with increasing metallicity.
Importantly, the N abundance on the surface can change due to 
the mixing of material in stars after the stars leave the main sequence. 
Thus, {\it the trend in the observed [N/Fe] ratio should be carefully compared with the model predictions}.

Since rotating massive stars produce a large amount of nitrogen at low metallicities
than non-rotating stars,
comparisons between models A and B show the impact of stellar rotation 
on the N abundance derived for Sculptor and Sextans dSphs. 
The [N/Fe] ratio of model A (orange curves) is higher 
than that of model B (green curves) at a given metallicity. 
The [N/Fe] ratio predicted by model A for Sculptor dSph seems 
roughly consistent with the average of the ratios of stars in the dSph 
corrected for the change in the ratio due to the mixing 
by stellar evolution models (grey dots in upper panel of 
Fig.~\ref{fig:ratio2}, and see the caption for the correction),
which suggests that the discrepancy between the observed [N/Fe] ratio
of the stars and the model prediction is likely due to
the internal mixing after leaving the main sequence.

With regard to the astrophysical sites, 
in addition to rotating massive stars, CNO elements can be produced 
during the stage of the hot bottom burning and the third dredge-up 
in intermediate mass stars 
\citep[e.g.][for a review]{1983ARA&A..21..271I}. 
The N yields for rotating massive stars also depend on stellar evolution models 
and nucleosynthesis calculations 
(see discussion in Section~3.4.2 in \citealt{2018MNRAS.476.3432P}).
The uncertainty associated with the measurement of the N abundance and 
the model does not allow us to conclude the sources and the contribution 
to the N abundance in the dSphs at this point.

\footnotetext{We have inferred the [N/Fe] ratio of individual red giants before the stars have experienced first dredge-up and extra mixing. Specifically, by referring to surface gravity, observed [C/Fe] and [N/Fe] of each star, the increase of the [N/Fe] ratio is very roughly estimated based on the predictions from stellar evolution models at [Fe/H]~$= -2.3$ shown in \citet{2014ApJ...797...21P}, where thermohaline mixing is taken into account to explain extra mixing. The modification of the surface abundance of low-mass red giant stars is likely explained mainly by thermohaline mixing \citep[e.g.][]{2007A&A...467L..15C}, and other mechanisms, such as magnetic fields and rotation, might cause the mixing. Also, observed stellar abundances suggest that the impact of extra mixing depends on metallicity \citep[e.g.][]{2019A&A...621A..24L,2019ApJ...872..137S}. The estimated initial [N/Fe] in the upper panel of Fig.~\ref{fig:ratio2} is shown to suggest the possible impact of internal mixing after the stars have left the main sequence, and thus the panel should be interpreted with caution.}

The elements heavier than zinc are synthesized through different processes.  
Although candidates of the intermediate neutron capture process can contribute to the chemical enrichment in dSphs \citep{2020A&A...634A..84S}, 
in this study, we focus on the s- and the r-processes that can be 
responsible to the production of most of the elements heavier than 
the Fe-peak. 

With regard to stellar abundance ratios, the abundance of trans-iron elements 
of stars in dwarf galaxies has been
measured, however, because of the low abundance in the stellar atmosphere, 
the quantity of the observational data is not always large. 
In particular, the chemical abundance of stars of low abundances may have not
been measured. 
When we will obtain the abundance of a large number of stars with next-generation
instruments and telescopes, 
our interpretations in this study might have to be revised.

\begin{figure*}
\centering
\includegraphics[scale=1.2]{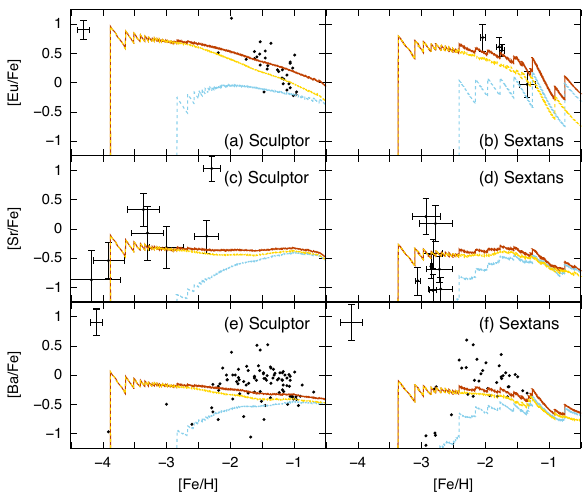}
\caption{The evolution of the [Eu/Fe], [Sr/Fe] and [Ba/Fe] ratios in Sculptor and Sextans dSphs (top, middle and bottom panels, respectively) as a function of metallicity.
Black dots are stellar abundance ratios.
The observational data are taken from \citet{2020A&A...641A.127R}.  
In panels (a), (e) and (f), the typical uncertainty about the stellar abundance ratios is shown at the upper left. 
To compare the abundance ratio of a larger number of stars with those predicted by models, 
observational data about the [Sr/Fe] ratio of Sextans stars are also gathered from 
\citet{2020A&A...636A.111A}  
and \citet{2022MNRAS.509.3626M}.  
When the abundance of a single star has been repeatedly measured in these studies, the abundance based on the latest data or with the small uncertainty is included.
Orange, light blue and yellow curves are the ratios predicted by models A, C and D, respectively. The assumptions about the relative contribution of the r-process sites in the model are described in Sec.~\ref{sec:model-2-1} and 
Appendix~\ref{app:2}.
}
\label{fig:ratio3}
\end{figure*}

Figure~\ref{fig:ratio3} shows the evolution of abundance ratios of Eu, 
Sr and Ba to Fe in Sculptor and Sextans dSphs. 
Eu is a representative element that is mainly 
($>$ 95~\% of the solar Eu abundance, \citealt{2020MNRAS.491.1832P})  
produced by the r-process, so we firstly focus on the [Eu/Fe] ratio to discuss 
the contribution of the r-process sites.

Akin to the [${\rm \alpha}$/Fe] ratio, the [Eu/Fe] ratio of stars in 
Sculptor dSph appears to decrease with metallicity, which can be due to
the ejection from SNe\,Ia \citep{2020A&A...641A.127R}. 
The ratio predicted by model A (solid orange curve) for Sculptor dSph 
shows a decreasing trend.
By comparing the [Eu/Fe] ratios of models C and D (dotted light blue and yellow curves),
we see that collapsars contribute to the 
chemical enrichment through the r-process greater than NSMs at lower metallicities. 
In particular, at the lowest metallicity, Eu is mainly produced by collapsars.
Since binary neutron stars reach the explosive event after the formation
of neutron stars, NSMs tend to contribute to the chemical enrichment at later time.
We note that the dispersion of the abundance ratio of stars at low metallicities
can also be explained by NSMs when a galaxy is assumed to be formed
through mergers of low-mass systems 
\citep[][for the case of the Galactic halo]{2015ApJ...804L..35I},
and the results have to be carefully interpreted.

Sr and Ba are the representative elements 
at the first and second peaks of the s-process, respectively.
As well as the s-process, these elements are produced through 
the r-process. 
Similar to the case of the [Eu/Fe] ratio, the contribution of collapsars 
is large relative to that of NSMs at the 
lowest metallicities (Figs.~\ref{fig:ratio3}c -- f).

\begin{figure*}
\centering
\includegraphics[scale=1.2]{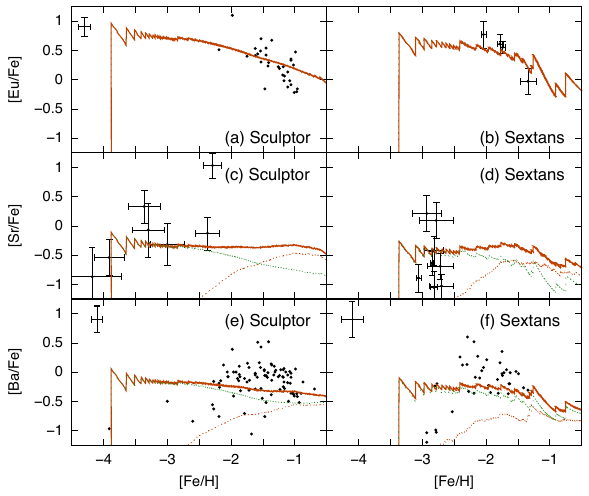}
\caption{Same as Fig.~\ref{fig:ratio3}, but the contribution of rotating
massive stars through the weak s-process and the contribution of the r-process
candidates are compared. The references of the observational data are
same as those in Fig.~\ref{fig:ratio3}.
Orange solid curves, green dotted curves and orange dotted curves show
the ratios predicted by models A, B and E, respectively.
}
\label{fig:ratio4}
\end{figure*}

Then, we investigate the contribution of rotating massive stars 
to the enrichment of trans-iron elements by the weak s-process and 
the contribution of the r-process candidates.
In Figures~\ref{fig:ratio4}a and b, Eu is mainly produced by the r-process.
Comparisons between the [Sr/Fe] ratio predicted by model A 
(solid orange curves in Fig.~\ref{fig:ratio4}) with that of model B 
(dotted green curves) show that rotating massive stars impact the Sr 
abundance at [Fe/H]~$\gtrsim -2$.
At lower metallicities, the abundance of heavy nuclei that become the seed of 
the neutron capture is low, and this might result in the small production 
of the elements through the weak s-process. 
Also, the [Sr/Fe] ratio predicted by model E (dotted orange curves) is much lower than those of
stars in Sculptor and Sextans dSphs,
which might suggest that the r-process sites contribute to the production of Sr.
These trends in the [Sr/Fe] ratio are also seen in the [Ba/Fe] ratio.

However, in dwarf galaxies of smaller masses, such as ultra-faint dwarf galaxies,
the chemical evolution proceeds slowly because of the low star formation rate,
and thus the abundance ratios of part of metal-poor stars might be
explained by the contribution of rotating massive stars
(Appendix~\ref{sec:app-uFd}).
When a dwarf galaxy is formed through mergers of lower-mass systems,
rotating massive stars can increase the abundance of Sr and Ba
through the weak-s process and create part of the dispersion of the
abundance ratio of the merged galaxy 
(Appendix~\ref{sec:app-bb}, online material).

We also point out that the results on the contribution of rotating massive stars
through the s-process depend on the evolution of massive stars.
For instance, according to the yields for a star of 
${\rm v_{ini}=150~km\,s^{-1}}$ at [Fe/H]~$=-2$, a single star of 
${\rm 30~M_{\odot}}$ releases ${\rm 1.3\times10^{-9}~M_{\odot}}$ of ${\rm {}^{88}Sr}$ through stellar winds when the star collapses into a black hole.
In the meanwhile, when the star evolves into a CCSN, in total
${\rm 1.4\times10^{-6}~M_{\odot}}$ of ${\rm {}^{88}Sr}$ is released. Thus,
if stars of $M > 25~{\rm M_{\odot}}$ evolve into CCSNe, 
rotating massive stars might contribute more to the production of 
trans-iron elements.

With regard to the sources of heavy elements, part of stars in mass range
of ${\rm 8 - 10~M_{\odot}}$ that evolve into ECSNe can produce light trans-iron 
elements through the weak r-process.
Also, rotating massive stars are expected to produce trans-iron nuclei through the s-,
the r- and the p-processes \citep[see discussion in][]{2022A&A...661A..86C}.
In addition, in low-metallicity environments where the neutron-to-seed ratio
can get high, nuclides at the second and third peaks of the s-process, such as
Ba and Pb, are expected to be more synthesized \citep[e.g.][]{1990A&A...234..211P}.
While our results suggest a minor contribution of rotating massive stars 
through the s-process to the
chemical enrichment with regard to the r-process sites at the lowest metallicities, the significance of the production of heavy nuclides in the environments 
of high neutron-to-seed ratios is still unclear.
For the r-process, the chemical yields for the candidates  
depend on the models.
Providing chemical yields at low metallicities and the uncertainty
associated with the nuclear data helps to understand the contribution of 
different sources of elements to the chemical enrichment.

\subsubsection{The impact of rare events on the abundance ratio}
\label{sec:4-2-2}
The chemical evolution progresses as individual events release 
the material.
When rare events significantly influence the chemical abundance,
the impact is expected to appear as the oscillation
in the abundance ratio. 

In Fig.~\ref{fig:ratio1}, 
the [Ca/Fe] ratio predicted by the model for Sculptor and Sextans dSphs does not 
show the significant oscillation at [Fe/H]~$\gtrsim-4$.
While both massive stars and SNe\,Ia can produce Ca and Fe,
Ca is mainly produced by massive stars 
and a large amount of iron is synthesized by SNe\,Ia in the model.
Thus, 
if massive stars and SNe\,Ia are rare enough, the [Ca/Fe] ratio is
expected to increase and decrease repeatedly every time these events appear.
As shown in Fig.~\ref{fig:nombre}, massive stars and SNe\,Ia are rare at the 
early time of the evolution of these dSphs, but as the star formation 
activity gets close to the peak, the number of these events in a time-step
increases and individual events give only weak impact on the [Ca/Fe] ratio.
In galaxies of mass as low as ${\rm M_*\sim10^{4-5}~M_{\odot}}$, massive stars 
and SNIa
are more rare and the impact on the [Ca/Fe] ratio might be seen,
as discussed in Appendix~\ref{sec:app-h}.

For the trans-iron elements, the impact of rare events is seen in the
abundance ratios.
In Fig.~\ref{fig:ratio3},
the [Eu/Fe] ratio predicted by models oscillates due to 
the r-process events. 
In Sculptor dSph, the impact of individual events (mainly collapsars) 
is seen at 
[Fe/H]~$\lesssim -3$. 
The oscillation in the [Eu/Fe]
ratio reduces with increasing metallicity, 
because the r-process events occur sufficiently to 
make the impact of individual events weak and that the abundance of Eu in the
galaxy gradually increases. 
Similarly, the [Eu/Fe] ratio in Sextans dSph oscillates due to collapsars
at the early time and the amplitude gets small
as the metallicity increases. 
The [Eu/Fe] ratio in Sextans dSph is influenced 
by the r-process events at higher metallicities.
The impact of NSMs on the [Eu/Fe] ratio in Sextans dSph appears as 
oscillation more clearly than that in Sculptor dSph
due to the small number of the r-process events.
Since NSMs occur at certain time intervals, the impact of these events clearly
appears as the oscillation in the [Eu/Fe] ratio predicted by model C.
The impact of the r-process sites depends on the yields and 
the relative contribution.
However, the oscillation in abundance ratios predicted by the models
for Sextans dSph might be partly because of the galaxy mass  
smaller than Sculptor dSph and the smaller number of the astrophysical events 
in the dSph.

In Fig.~\ref{fig:ratio3}c, the [Sr/Fe] ratio of stars in Sculptor dSph shows 
the scatter that seems to be greater than the uncertainty 
about the measurement at [Fe/H]~$\lesssim -2.5$.
The ratio predicted by models A and D for this dSph oscillates at 
low metallicities mainly due to 
individual events of collapsars, and
thus rare events might be a cause of the dispersion in the stellar abundance.
When we see the [Ba/Fe] ratio of stars in Sculptor dSph (Fig.~\ref{fig:ratio3}e),
there is the dispersion at [Fe/H]~$\gtrsim -2.5$.
In this metallicity range, the [Ba/Fe] ratio predicted by the models
evolves almost smooth, because the number of events that produce Ba
gets large.
Fig.~\ref{fig:ratio3}f also shows that there seems to be the dispersion 
in the [Ba/Fe] ratio of stars in Sextans dSph at [Fe/H]~$\sim -2$.
In contrast to Sculptor dSph, the ratio predicted by the models for Sextans dSph
oscillates even at metallicity higher than [Fe/H]~$\sim -2.5$, because
of the rarity of the r-process events,
and thus part of the dispersion at [Fe/H]~$\gtrsim -2.5$ might come from
the rare r-process events (both NSMs and collapsars).

The comparisons between the impact of rare events on the abundance ratio
predicted for Sculptor and Sextans dSphs 
show that the abundance ratio of a lower-mass galaxy might oscillate
more frequently.
In addition, the amplitude of the oscillation seems to be larger for a galaxy
of a smaller mass, as seen in the abundance ratio predicted by models
A and C for Sextans dSph.
Thus, the abundance ratio of a lower-mass
galaxy might be more influenced by rare events
(see Appendix~\ref{sec:app-uFd} for the case of ultra-faint dwarf galaxies).

In the meanwhile, the oscillation in the abundance ratio predicted 
by the model seems small relative to the dispersion 
seen in the abundance ratio of stars, 
which suggests that the dispersion in the abundance ratio of stars may 
come from different processes.
Also, in the framework of the ${\rm \Lambda}$-CDM model, more massive galaxies
are formed through mergers of systems of smaller masses.
Although there might be the discrepancy between observations and predictions 
based on the ${\rm \Lambda}$-CDM model in a small scale 
\citep[][for a review]{2017ARA&A..55..343B}, 
pieces of observational evidence, such as the structure in dSphs found 
by photometries \citep[e.g.][]{2018MNRAS.480..251C} and 
the ${\rm \alpha}$-knees \citep[e.g.][]{2020A&A...641A.127R},
suggest that dSphs might have experienced mergers in the past.
Before merging into a galaxy, each building-block low-mass system is enriched
in heavy elements and has a unique chemical enrichment history, 
which is expected to be significantly influenced by rare events.
In particular, when rare events occur in building-block systems where
the chemical enrichment proceeds slow,
the impact of the rare events can be reflected in the dispersion of the
abundance ratios at low metallicities \citep[e.g.][]{2015ApJ...804L..35I}.
Therefore, if a dSph is comprised of lower-mass systems, the large impact of 
individual rare events on the abundance ratio is expected to appear.
We discuss the simplest case in Appendix~\ref{sec:app-bb} (online material).

\section{Conclusions}

We investigate the chemical evolution of two dSphs around the Milky Way
(Sculptor and Sextans dSphs) 
using a chemical evolution model where the rarity of the source of 
elements is introduced.
We focus on the impact of 
rare events and the contribution of astrophysical sites
to the chemical enrichment.
In the model, the occurrence of each source of elements (massive stars
of different rotating velocities, LIMS, SNe\,Ia and the r-process candidates)
follows the star formation history of the dSphs derived from the colour-magnitude diagram
and the chemical evolution progresses through the ejection of elements 
from individual events.
The rates of the gas accretion and the outflow are determined
based on comparisons between the metallicity distributions of the 
dSphs and those predicted by the model.

We compare abundance ratios of stars in the dSphs to those predicted by  
the model. 
Despite the simplicity, the model captures the average of the [Ca/Fe] ratio
of the stars in the dSphs.

We attempt to assess the contribution of rotating massive stars to the
N abundance in the dSphs. The model that includes rotating massive
stars predicts low [N/Fe] ratios relative to those of red giant stars.
The disagreement is most probably due to the change of 
the N abundance of the red giant stars
resulting from internal mixing at an earlier stage of their evolution.
It is also possible that 
massive AGB stars with the hot bottom burning, 
besides rotating massive stars, contribute to 
the enrichment of nitrogen.

We also investigate the contribution of rotating massive stars through 
the weak s-process and the contribution of the r-process sites 
to the chemical evolution.
The contribution of rotating massive stars through the weak s-process 
is seen at [Fe/H]~$\gtrsim -2$, and at lower metallicities, 
the contribution of the r-process sites seems large.
The small contribution of rotating massive stars through the s-process 
at the lowest metallicities
is supposed to be due to the small abundance of the seed nuclei. 
However, in lower-mass dwarf galaxies, such as ultra-faint dwarf galaxies,
massive stars of different initial rotating velocities might more impact
the chemical evolution.
If a dwarf galaxy is formed through mergers of less massive systems,
the contribution of rotating massive stars to the chemical enrichment
in the lower-mass systems may create part of the dispersion
of the abundance ratios of the merged dwarf galaxy.
This issue still remains open to discuss because of 
the uncertainty about the stellar evolution and the yields.
Further investigations on chemical yields 
and observational information, such as the chemical abundance of metal-poor stars and the rates of the r-process events, will
bring understanding about the sources of elements and the
chemical enrichment in the low metallicity environments.

With regard to the impact of rare events on the chemical abundance,
the [Ca/Fe] ratio predicted for Sculptor and Sextans dSphs evolves
almost smooth at [Fe/H]~$\gtrsim -4$, since massive stars and SNe\,Ia 
frequently appear.
On the other hand, for trans-iron elements, 
we see the oscillation in the abundance ratios predicted by the
model for the dSphs at the lowest metallicities 
through the r-process.
The amplitude of the oscillation mainly generated by collapsars at the
lowest metallicities gets small as the metallicity increases.
While the abundance ratios in Sculptor dSph get almost smooth at 
[Fe/H]~$\gtrsim -3$, we observe the impact of NSMs on the abundance ratios
in Sextans dSph even at higher metallicities.
Since the galaxy mass of Sextans dSph is smaller than that of Sculptor dSph,
the occurrence of the r-process events is also generally small, 
which might result in
the oscillation of the abundance ratios at higher metallicities.
This might suggest that rare events play an important role on
the abundance ratio when a galaxy is formed from systems of smaller masses.


\section*{Acknowledgements}
NF is grateful for the opportunity of research activities at IAP in the winter of 2019 -- 2020 partly supported by The Graduate University for Advanced Studies, SOKENDAI. 
Numerical calculations in this study have been carried out with PC cluster at Center for Computational Astrophysics, NAOJ.

\section*{Data Availability}
No new data were generated or analysed in support of this research.
 



\bibliographystyle{mnras}




\appendix

\section{The variation of the number of events with time-step}
\label{sec:app-tstep}
\begin{figure*}
\centering
\includegraphics[scale=0.8]{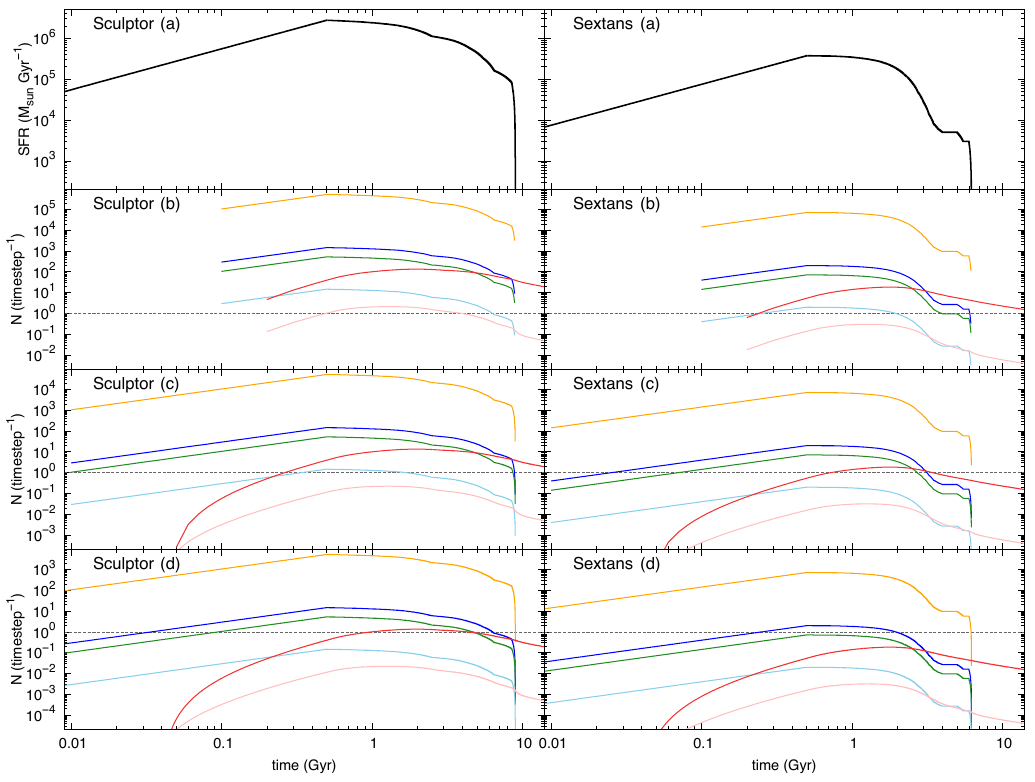}
\caption{Same as Figs.~\ref{fig:nombre}a and b, but different sizes of time-step
are adopted.
(a) The star formation history of Sculptor and Sextans dSphs referred from
\citet{2012A&A...539A.103D}  
and \citet{2018MNRAS.476...71B}, 
respectively.
(b) The time variation of the number of events in each time-step ($N_{\rm event}$)
when ${\rm \Delta t=10^{-1}}$~Gyr is adopted.
(c and d) Same as panels (b), but the cases of ${\rm \Delta t=10^{-2}}$ and $10^{-3}$~Gyr, respectively.
Panels (a) and (d) have been already shown in Fig.~\ref{fig:nombre},
but we include them in this figure for reference.
The colour of curves corresponds to the source of the chemical evolution,
as described in the caption of Fig.~\ref{fig:nombre}.
Horizontal dotted lines show $N_{\rm event}=1$.
}
\label{fig:app-tstep}
\end{figure*}

As seen in equation \ref{eq:nombre}, the number of events in a time-step
($N_{\rm event}$) depends on the time-step $\Delta t$.
Since we consider an event to be rare when $N_{\rm event} < 1$,
whether the event is rare also depends on the size of time-step.
Figure~\ref{fig:app-tstep} shows the time variation
of $N_{\rm event}$ for Sculptor and Sextans dSphs
when time-steps
$\Delta t = 10^{-1}, 10^{-2}$ and $10^{-3}$~Gyr are adopted.

Generally, once the star formation history is fixed, 
the number of events in a given time-step gets smaller when a smaller size of 
${\rm \Delta t}$ is adopted.
For instance, collapsars (light blue curve)
are always rare in Sculptor dSph when the time-step is set to be
${\rm \Delta t = 10^{-3}}$~Gyr. 
In the meanwhile, they are not considered to be rare when 
${\rm \Delta t = 10^{-1}}$~Gyr is adopted except for the late time of the evolution.

As stated in the main text, in this study we adopt ${\rm \Delta t = 10^{-3}}$~Gyr
by taking into account the lifetime of massive stars and the precision
in the calculations.

\section{The relative contribution of the astrophysical sites of the r-process}
\label{app:2}

As described in Section~\ref{sec:model-2-1}, the r-process sites, the yields and
the contribution to the chemical enrichment are under debate.
To examine the contribution of the r-process sites, 
we firstly discuss the variation of abundance ratios in the Milky Way with the relative contribution between NSM and collapsar.

The Galactic disc has been formed through physical processes,
such as the accretion of gas and the stellar radial migration.
To see the contribution of each astrophysical site to the chemical enrichment,
we simplify the evolution of the Galactic disc and 
compare the observables of the whole (thin and thick) disc to the
one-zone infall model.

The one-zone model is prepared along \citet{2008EAS....32..311P} and for
the chemical enrichment, \citet{2015A&A...580A.126K} 
and \citet{2018MNRAS.476.3432P} are referred to. 
We assume that the Galactic disc has been formed through the accretion of metal-free gas.
In the model, all of gas is in the gas reservoir of mass $M_{\rm g,h}$ at $t=0$ (Gyr). 
The masses of stars and gas in the galaxy are $M_{\rm s}$ and $M_{\rm g}$.
The total mass of the system is $M=M_{\rm g}+M_{\rm s}+M_{\rm g,h}$.
The mass fractions are defined as $m_{\rm g,h}=M_{\rm g,h}/M$, $m_{\rm g}=M_{\rm g}/M$ 
and $m_{\rm s}=M_{\rm s}/M$, respectively.
The gas accretes on to a galaxy at the infall rate $f(t)$ that is assumed to be proportional
to the mass of gas in the reservoir ;
\begin{equation}
f(t) = {\rm k_{in}} \, m_{\rm g,h}(t),
\end{equation}
where ${\rm k_{in}}$ is a constant that regulates the time-scale of the 
accretion of gas. We set ${\rm k_{in}=0.1~Gyr^{-1}}$ as a typical value.
In the galaxy, the gas turns into stars at the star formation rate $\psi(t)$
\citep[][]{1959ApJ...129..243S} ;
\begin{equation}
\psi(t) = {\rm k_{SF}} \, m^{\rm N}_{\rm g}(t),
\end{equation}
where ${\rm k_{SF}}$ is a coefficient called as the star formation 
efficiency\footnote{The definition of the star formation efficiency depends on 
studies.
We refer to the coefficient ${\rm k_{sf}}$ of the star formation 
rate as the star formation efficiency.}.
The efficiency is determined as the metallicity distribution 
\citep[][]{2014A&A...562A..71B} and the present-day 
gas fraction of the Galactic disc (\citealt{2015A&A...580A.126K},  
references therein) are roughly explained by the model.
The index ${\rm N}$ is set to be ${\rm N=1.4}$.
Regarding the IMF, \citet{2002ASPC..285...86K}  
IMF is referred.
For the stellar lifetime, theoretical values of the case of ${\rm Z}={\rm Z}_{\odot}$ in \citet{1992A&AS...96..269S}  
are adopted. 

In this study, we assume that physical quantities related to the 
chemical enrichment are not varied among the Milky Way and dwarf galaxies.
Thus, the assumptions about the sources of the chemical enrichment are 
almost identical between the model for the Galactic disc and that for dSphs
(Sec.~\ref{sec:3-2}). 
Since the Milky Way is a relatively massive (about ${\rm 5\times10^{10}~M_{\odot}}$ in stars), 
we do not adopt the concept of the rarity\footnote{Note, however, 
that the Milky Way has been formed through mergers and accretions of 
other systems. If the building-block systems have low 
masses and low star formation rates, the astrophysical events can be rare 
in these systems.}. 
SNIa and NSM occur certain time after the formation of
white dwarfs and neutron stars, respectively. 
The rates at time $t$ are calculated by 
\begin{equation}
R(t) = \int_0^t \Psi(t') \,DTD(t - t') \,{\rm d}t',
\end{equation}
where $t'$ represents the time of the star formation.
The DTDs are referred from \citet[][references therein]{2017ApJ...848...25M} 
and \citet{2005A&A...441.1055G}  
for SNIa (the details are described in Sec.~\ref{sec:3-2}) and 
\citet{2019MNRAS.487.4847B} 
for NSM, respectively.
We adopt the yields for the r-process sites as assumed in Sec.~\ref{sec:model-2-1}.

\begin{figure}
\centering
\includegraphics[scale=1.2]{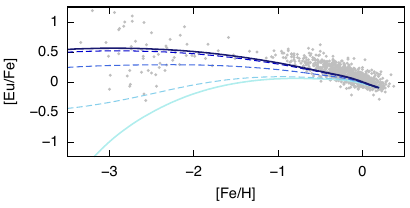}
\caption{The variation of the [Eu/Fe] ratio predicted by the model 
for the Galactic disc 
with the relative contribution of the r-process sites (${\rm A_r}$ 
in the yields for the r-process sites). 
The model includes massive stars of different initial rotating velocities, 
LIMS, SNe\,Ia and the r-process 
candidates in the source of the chemical enrichment.
A curve in deeper blue corresponds to a case of a larger contribution of 
collapsars to the solar Eu abundance. 
The solid curve in the deepest blue shows the ratio when collapsar alone 
is assumed to be the dominant site of the r-process (${\rm A_r=1}$), 
while the solid curve in the lightest blue is the case where NSM alone is assumed 
to be the dominant r-process site (${\rm A_r=0}$). 
Dotted curves show the cases where both collapsar and NSM contribute 
to the solar Eu (${\rm A_r=0.1, 0.5}$ and $0.9$). 
Grey dots are the ratio of stars in the Milky Way. 
The observational data are taken 
from \citet{2005A&A...439..129B,2013A&A...552A.128M,2016A&A...586A..49B},
and \citet{2021A&A...649A.126T}. 
}
\label{fig:r-yield}
\end{figure}

In Figure~\ref{fig:r-yield}, we compare the [Eu/Fe] ratio of stars in the Milky Way with those predicted by the model 
that includes massive stars of different initial rotating velocities, 
LIMS, SNe\,Ia and the r-process sites in the source of elements.
When the dominant r-process site is assumed to be NSM (${\rm A_r}$ in 
the yields ${\rm A_r=0}$, a solid curve in the lightest blue), 
the [Eu/Fe] ratio predicted by the model is lower than those of the Milky Way stars. 
This is because NSMs occur certain time after the formation of neutron
stars. The delay time distributes in the range of $\sim0.03 - 10^5$~Gyr
\citep{2019MNRAS.487.4847B}.  

On the other hand, when collapsar
is the dominant site of the r-process (${\rm A_r=1}$, a solid curve in the deepest blue), 
the [Eu/Fe] ratio decreases with increasing metallicity.
Since massive stars have short lifetimes (typically about $10^7$~years), 
collapsars can contribute
to the enrichment of heavy elements even at the early time of the
Milky Way \citep[e.g.][]{2015MNRAS.452.1970W}. 

Since both NSM and collapsar are not
excluded from the candidates, 
we assume that both are the main sites of the r-process. 
Since collapsars can produce 
heavy elements at low metallicities, a larger contribution of collapsars 
(a larger ${\rm A_r}$)
results in a higher [Eu/Fe] ratio at a given metallicity.

When NSMs produce a large part of the solar Eu, 
the predicted [Eu/Fe] ratio tends to be lower than the ratio of the Milky Way stars.
In the meanwhile, there is the observational evidence for the nucleosynthesis 
through the r-process 
at NSM. 
Also, we cannot exclude the possibility that stars of low Eu abundances will be 
found and the average of the [Eu/Fe] ratio of stars gets lower than that 
of stars in Fig.~\ref{fig:r-yield}. 
Thus, in this study, the relative contribution 
is allowed to be in the range of ${\rm 0 \leq A_r \leq 1}$.

Then, we investigate the contribution of the r-process sites to the
chemical evolution in Sculptor and Sextans dSphs.
It is a matter of debate whether the relative contribution of the
r-process sites to the chemical enrichment varies among different systems,
while the assessment for stellar systems in the Galactic halo suggests that
the contribution of special classes of CCSNe to the enrichment of Eu 
might be different among the systems \citep{2022ApJ...926L..36N}.
Also, since the occurrence of the r-process events is low, 
the r-process events might occur in only a part of low-mass systems
of stellar mass as small as ${\rm M_*\sim10^4~M_{\odot}}$
and enrich the systems in heavy elements through a single event
\citep[e.g.][]{2016Natur.531..610J}.
Thus, the variation of the relative contribution of the r-process sites
to the chemical enrichment among low-mass galaxies might not be excluded. 
In this study, we allow the relative contribution to vary among 
dwarf galaxies and roughly set the value of ${\rm A_r}$ for each dSph
based on the abundance ratios.

\begin{figure}
\centering
\includegraphics[scale=0.85]{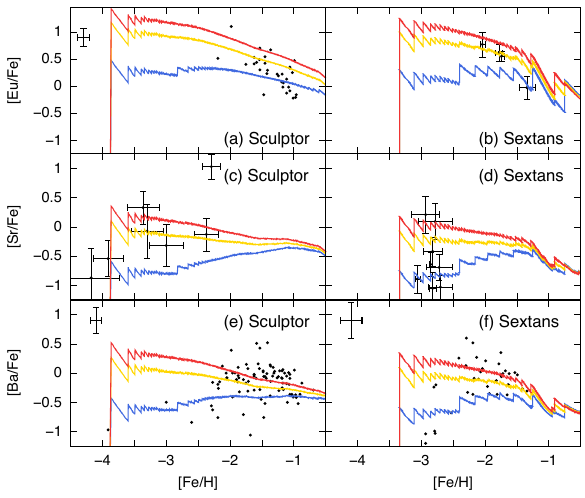}
\caption{(a) and (b) The variation of the [Eu/Fe] ratio predicted by model A (the sources of the chemical enrichment are summarized in Table.~\ref{tab:model}) for Sculptor and Sextans dSphs with the relative contribution of the r-process sites (${\rm A_r}$). Blue, yellow and red curves correspond to cases of different contributions of the r-process sites (${\rm A_r=0.1, 0.5}$ and $0.9$, respectively). (c -- f) Same as panels (a) and (b), but the cases of the [Sr/Fe] and [Ba/Fe] ratios. In all panels, black dots show the ratios of stars in the dSphs. The references of the observational data are same as those in Fig.~\ref{fig:ratio3}.
}
\label{fig:r-yield2}
\end{figure}

In Figure~\ref{fig:r-yield2}, we show the variation of the [Eu/Fe], [Sr/Fe] and [Ba/Fe] ratios 
in Sculptor and Sextans dSphs predicted
by model A with different relative contributions of the r-process sites
(${\rm A_r=0.1, 0.5}$ and $0.9$).
When we assume a larger contribution from collapsars 
(a larger ${\rm A_r}$), the [Eu/Fe] ratio is higher at a given
metallicity, as seen in the case of the Milky Way.
This is because Eu is produced by collapsars
 at the early stage of the evolution of the dSphs. 
A similar trend is seen in the [Sr/Fe] and [Ba/Fe] ratios.

In the case of Sculptor dSph (left panels in Fig.~\ref{fig:r-yield2}), 
when ${\rm A_r}$ is assumed to be ${\rm A_r = 0.1}$,
the predicted [Eu/Fe] ratio seems lower than the average of the stars.
Also, the [Sr/Fe] and [Ba/Fe] ratios of the model are generally 
low relative to those of stars. 
When ${\rm A_r}$ is assumed to be ${\rm A_r = 0.9}$,
the [Eu/Fe] ratio is higher than the average of stars.
We set the relative contribution to be ${\rm A_r = 0.3}$ for Sculptor dSph.

In the case of Sextans dSph (right panels), 
when we assume ${\rm A_r = 0.9}$,
the [Sr/Fe] ratio is higher than the average of stars in the dSph.
Also, the [Eu/Fe] ratio of stars of low Eu abundances may not have been measured.
We assume ${\rm A_r = 0.3}$ for Sextans dSph.

While we adopt ${\rm A_r = 0.3}$ to the dSphs, we stress
that the variation of the relative contribution of the r-process sites
among different systems needs to be further investigated based on the
yields for the r-process sites and plenty of stellar abundance.

\section{The impact of individual events on abundance ratios in dwarf galaxies of different galaxy masses}
\label{sec:app-h}
To investigate the variation of the impact of rare events on 
the chemical abundance with galaxy mass, we compare
abundance ratios of hypothetical dSphs of different galaxy stellar masses.

We consider dSphs a, b, c and d of stellar mass ${\rm M_*=10^6, 5\times10^5, 10^5}$ and 
${\rm 10^4~M_{\odot}}$, respectively.
In each galaxy, stars are assumed to be formed at the early stage 
of the evolution during ${\rm T_{SF}=2.5}$~Gyr (arbitrarily
set based on the star formation history of Sextans dSph) at constant star formation
rates ${\rm \Psi=M_*/T_{SF}}$.
These galaxies are on the stellar mass -- stellar metallicity relation
\citep[][]{2013ApJ...779..102K}.
The properties of the hypothetical dSphs are summarized in Table~\ref{tab:app-h}.

As in the main text, these galaxies are assumed to be the system that evolves through
the accretion of gas and the outflow.
For the purpose of comparing the impact of rare events on the chemical abundance,
we set the time-scale of the gas accretion ${\rm \alpha}$ to be ${\rm \alpha=1.5~Gyr^{-1}}$
for these hypothetical dSphs and the gas accretion is assumed to stop at $t=2.5$~Gyr.
The representative efficiency of the outflow ${\rm k}$ is roughly determined for each dSph as the metallicity at the peak
of metallicity distribution derived by the model is consistent with the average metallicity
of the dSph.
In the case of dSph d, the metallicity distribution is discrete due to the small number of 
stars formed in the dSph.
Thus, the metallicity at the median of the distribution is compared to the average metallicity.
We assume that the hypothetical dSphs lose their gas through the star
formation and the outflow due to CCSNe.

\begin{table}
	\centering
        {\scriptsize
	\caption{Properties of hypothetical dSphs.}
	\label{tab:app-h}
	\begin{tabular}{lllll} 
		\hline
		 & a & b & c & d \\
		\hline
		galaxy mass in stars (${\rm M_{\odot}}$) & $10^6$ & $5\times10^5$ & $10^5$ & $10^4$  \\
		average metallicity ($\langle$[Fe/H]$\rangle$) & $-1.69$ & $-1.78$ & $-1.99$ & $-2.29$ \\
		star formation rate (${\rm M_{\odot}Gyr^{-1}}$)& $4\times10^5$ & $2\times10^5$ & $4\times10^4$ & $4\times10^3$ \\
		duration of star formation (Gyr) & $2.5$ & $2.5$ & $2.5$ & $2.5$  \\
                time-scale of gas accretion (${\rm \alpha~Gyr^{-1}}$) & $1.5$ & $1.5$ & $1.5$ & $1.5$ \\
		representative efficiency of outflow (${\rm k}$) & $3.0\times10^3$ & $3.5\times10^3$ & $5.0\times10^3$ & $8.0\times10^3$ \\
		\hline
	\end{tabular}
}
\end{table}

\begin{figure*}
\centering
\includegraphics[scale=0.6]{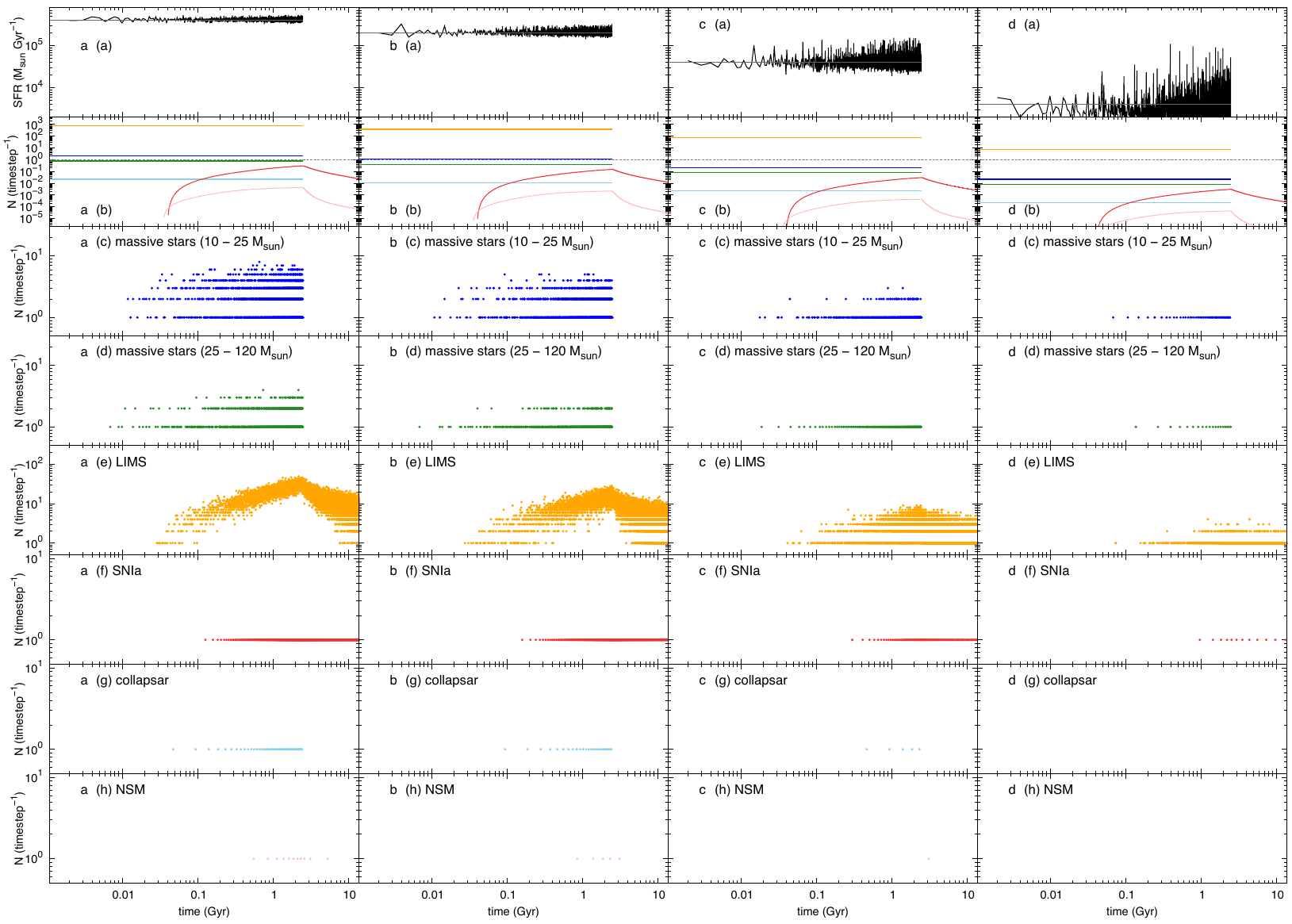}
\caption{Same as Fig.~\ref{fig:nombre}, but for hypothetical dSphs a, b, c and d
(the first, second, third and fourth columns, respectively).
Black curves in top panels show the modelled star formation histories
after the rarity is introduced. 
}
\label{fig:app-h1}
\end{figure*}

Similar to Fig.~\ref{fig:nombre} in the main text, Figure~\ref{fig:app-h1} shows the time variation of the star formation rate,
the number of events in a time-step before the rarity is introduced and the number of events that release the material in a
time-step for dSphs a, b, c and d.
Generally, the number of events that release the material in a time-step is smaller in a galaxy of a smaller mass.
Due to the low occurrence of the r-process events, 
neither collapsar nor NSM occurs in galaxy d that has the 
smallest mass among the four dSphs.
Galaxy d is regarded as the representative of low-mass dSphs that have not experienced the r-process events. 

\begin{figure}
\centering
\includegraphics[scale=0.85]{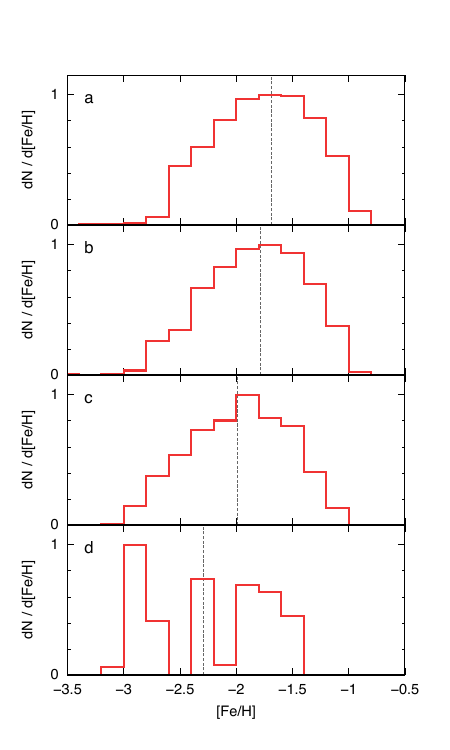}
\caption{Metallicity distributions of dSphs a, b, c and d. The vertical dotted lines show the
average metallicity of each galaxy.
}
\label{fig:app-h2}
\end{figure}

Adopting the number of events shown in Fig.~\ref{fig:app-h1}, we derive the chemical abundance for dSphs a, b, c and d.
Figure~\ref{fig:app-h2} shows the metallicity distributions. 
Since the dSphs are assumed to follow the stellar mass -- stellar metallicity relation,
a galaxy of a smaller mass has a lower average metallicity.

\begin{figure}
\centering
\includegraphics[scale=0.85]{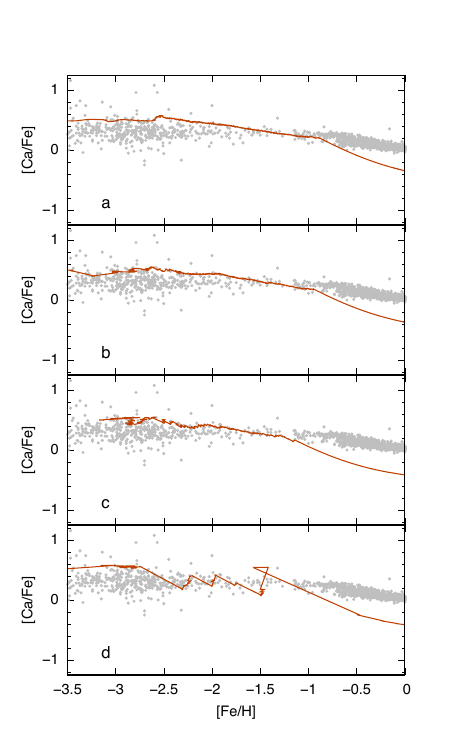}
\caption{The evolution of the [Ca/Fe] ratio in galaxies a, b, c and d predicted by 
model A (orange curves) is shown as a function of metallicity.
Stars in the Milky Way are shown in grey dots for the purpose of observing the impact of
individual events on the abundance ratio. The observational
data are taken from \citet{2000A&AS..141..491C,2005A&A...439..129B,2012A&A...545A..32A,2013ApJ...762...26Y}
and \citet{2014A&A...562A..71B}.
}
\label{fig:app-h3}
\end{figure}

The impact of individual events on abundance ratios might be more clearly seen in galaxies of lower masses.
Figure~\ref{fig:app-h3} shows the evolution of the [Ca/Fe] ratio in the dSphs 
as a function of metallicity.
As we have described in Sec.~\ref{sec:6}, 
while Ca and Fe are produced by massive stars and SNe\,Ia,
Ca is mainly produced by massive stars and a SNIa releases a large amount
of Fe in the model.
Thus, when massive stars and SNe\,Ia are rare, the [Ca/Fe] ratio is expected to
oscillate due to the rarity.
The [Ca/Fe] ratio of galaxies a and b is almost smooth
at [Fe/H]~$>-3.5$, since massive stars and SNe\,Ia frequently appear. 
When the galaxy mass is as small as that of dSph c, the weak oscillation is seen in the ratio at [Fe/H]~$\lesssim-2.5$, where the occurrence of 
massive stars and SNe\,Ia is low, as shown in Figs.~\ref{fig:app-h1}d and f.
In the case of galaxy d, massive stars and SNe\,Ia are more rare,
and the impact of individual events can be clearly seen in the [Ca/Fe] ratio.
Due to the ejection of iron from a SNIa,
the [Ca/Fe] ratio decreases when a SNIa occurs.

\begin{figure}
\centering
\includegraphics[scale=0.85]{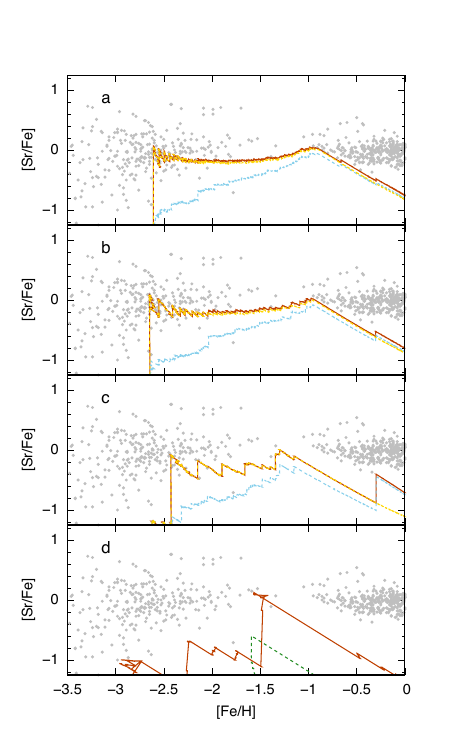}
\caption{The evolution of the [Sr/Fe] ratio in galaxies a, b, c and d predicted by the
model.
For galaxies a, b and c, the [Sr/Fe] ratio predicted by models A, C and D 
(orange, light blue and yellow curves, respectively) is shown as a function of metallicity.
In dSph d, no r-process event occurs throughout the evolution, 
and thus massive stars and LIMS contribute to the production of Sr 
through the s-process. 
Orange and green curves in the lowest panel show the ratio derived by
models E and F for dSph d, respectively.
The difference in the [Sr/Fe] ratio predicted by models E and F shows the
impact of rotating massive stars through the weak s-process on the ratio.
Stars in the Milky Way are shown in grey dots, as in Fig.~\ref{fig:app-h3}. 
The observational data are taken from \citet{2005A&A...439..129B,2008ApJ...681.1524L,2016A&A...586A..49B}
and \citet{2021A&A...649A.126T}.
}
\label{fig:app-h4}
\end{figure}

Figure~\ref{fig:app-h4} shows the [Sr/Fe] ratio predicted by models A, C and D listed in Table~\ref{tab:model} for dSphs a, b and c
as a function of metallicity. 
Regarding the r-process, the relative contribution between collapsar
and NSM is fixed to be ${\rm A_r=0.5}$ to see the impact of rare events on the
chemical abundance in dSphs of different masses.
For galaxy d, the evolution of the ratio predicted by models E and F is shown.
No r-process event occurs in galaxy d, and thus the s-process in rotating massive stars
and LIMS contribute to the production of Sr in this galaxy.

\begin{figure}
\centering
\includegraphics[scale=0.85]{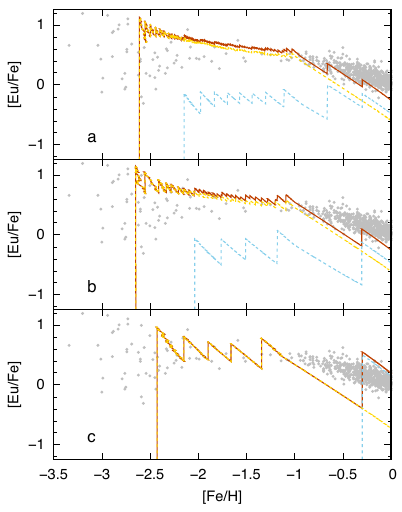}
\caption{Same as Fig.~\ref{fig:app-h4}, but the evolution of the [Eu/Fe] ratio.
Since no r-process event occurs in galaxy d and that Eu is mainly produced
by the r-process, the [Eu/Fe] ratio of dSph d is not shown.
Observational data about the Galactic stars are same as those shown in 
Fig.~\ref{fig:r-yield}.
}
\label{fig:app-h5}
\end{figure}

As is the case with Sculptor and Sextans dSphs, the oscillation in the [Sr/Fe] ratio
in galaxies a, b and c
at low metallicities is caused mainly by collapsars
through the r-process.
The amplitude of the oscillation in the ratio of dSph b at low metallicities seems not significantly different from that of dSph a. 
On the other hand, the impact of five collapsars and
one NSM 
is clearly seen in the [Sr/Fe] ratio of galaxy c.
This suggests that rare events can be part of the cause of the dispersion in abundance ratios
of lower-mass dSphs.
Similar trends are seen in the [Eu/Fe] ratios of the dSphs 
(Figure~\ref{fig:app-h5}).

In addition, we see the impact of rotating massive stars through the s-process
on the [Sr/Fe] ratio for dSph d.
In particular, the [Sr/Fe] ratio predicted by model E sharply increases
at [Fe/H]~$\sim -1.5$ due to the ejection of Sr by a rotating massive star.
When this massive star releases material, the amount of the interstellar gas 
in galaxy d is sufficiently small because of the star formation and the outflow.
Since the rotating massive star releases Sr into a small amount of gas,
the [Sr/Fe] ratio temporally increases to [Sr/Fe]~$\sim 0$.

Considering a more massive galaxies is formed through mergers of lower mass systems
in the framework of the cold dark matter model,
the rarity might play a role on part of the dispersion in the chemical abundance of
galaxies.

\section{The impact of individual events on abundance ratios in ultra-faint dwarf galaxies}
\label{sec:app-uFd}

To discuss the impact of rare events on the chemical enrichment 
in low-mass systems, 
we investigate the chemical evolution of dwarf galaxies of mass 
as small as those of ultra-faint dwarf galaxies (uFds). 
As done in the main text, we estimate the number of astrophysical sources 
of the chemical enrichment based on the observationally-derived 
star formation history, introduce the rarity and compare abundance ratios
predicted by the model with those of stars in uFds. 
The number of observed stars in individual uFds is not always large, 
and we focus on the case of hypothetical uFds to discuss 
general trends of abundance ratios.

Firstly, we estimate the number of each astrophysical source in a time-step. 
While the star formation history of uFds may be derived 
from sparse colour-magnitude diagrams, 
uFds are generally dominated by old and metal-poor stars
\citep[e.g.][]{2005ApJ...626L..85W,2006ApJ...643L.103Z,2007ApJ...654..897B,2008ApJ...688..245W},
and the majority of the stars are likely to have been formed at the early time 
(within the first one or two Gyrs) of the evolution
\citep[e.g.][]{2014ApJ...796...91B}. 
We consider the average star formation history of 
non-Magellanic satellites measured by \citet{2021ApJ...920L..19S}
to be a typical star formation history of uFds.

According to spectroscopic analyses, 
most of the stars in uFds show low abundance ratios of 
heavy trans-iron elements to iron, 
while a small number of uFds, such as Reticulum~II, 
include stars of high abundance ratios
\citep[e.g.][]{2016Natur.531..610J,2016AJ....151...82R}.
This is interpreted as r-process events appear 
in a small number of uFds and enrich the interstellar medium 
in heavy elements. 

To observe the impact of rare events on abundance ratios, 
we consider three cases. 
In the first case, no r-process event appears in an uFd (uFd i). 
In the second case, one collapsar occurs in an uFd at a certain time 
(uFds ii, iii and iv). 
In the third case, a NSM appears during the evolution 
(uFds v, vi and vii). 
All of the seven hypothetical uFds are assumed to have the identical 
stellar mass 
${\rm M_* = 5\times10^3~M_{\odot}}$ and star formation history, 
but the properties of individual stars 
formed in each galaxy and the time of the appearance of an r-process event 
(for uFds ii -- vii) are different.
The total number of events in uFds i -- vii and 
the time of appearance of r-process events are summarized in Table~\ref{tab:nom-uFd}.

\begin{table*}
  \centering
        {\scriptsize
          \caption{The number of astrophysical sources in unit mass (${\rm n_{event}}$) and the total number of astrophysical sources throughout the evolution in the seven hypothetical uFds. 
For massive stars and LIMS, the number of stars formed in the uFds is shown.
For the r-process events in the uFds, the time of the appearance is described in parentheses in Gyr.}
          \label{tab:nom-uFd}
              \begin{tabular}{lllllllll} 
                \hline
                 & ${\rm n_{event}}$ & i & ii & iii & iv & v & vi & vii \\
                \hline
                massive star (${\rm 10-25~M_{\odot}}$)& $5.3\times10^{-3}$ & \multicolumn{7}{c}{$42$} \\
                massive star (${\rm 25-120~M_{\odot}}$) & $1.9\times10^{-3}$ & \multicolumn{7}{c}{$15$} \\
                LIMS & $1.9$ & \multicolumn{7}{c}{$15079$} \\
                SNIa & $1.3\times10^{-3}$ & \multicolumn{7}{c}{$9$} \\
                NSM & $1.2\times10^{-5}$ & 0 & 0 & 0 & 0 & 1 (0.310) & 1 (1.483) & 1 (2.629) \\
                collapsar & $5.3\times10^{-5}$ & 0 & 1 (0.080) & 1 (0.323) & 1 (1.171) & 0 & 0 & 0 \\
                \hline
                \end{tabular}
}
\end{table*}

\begin{figure}
\centering
\includegraphics[scale=0.8]{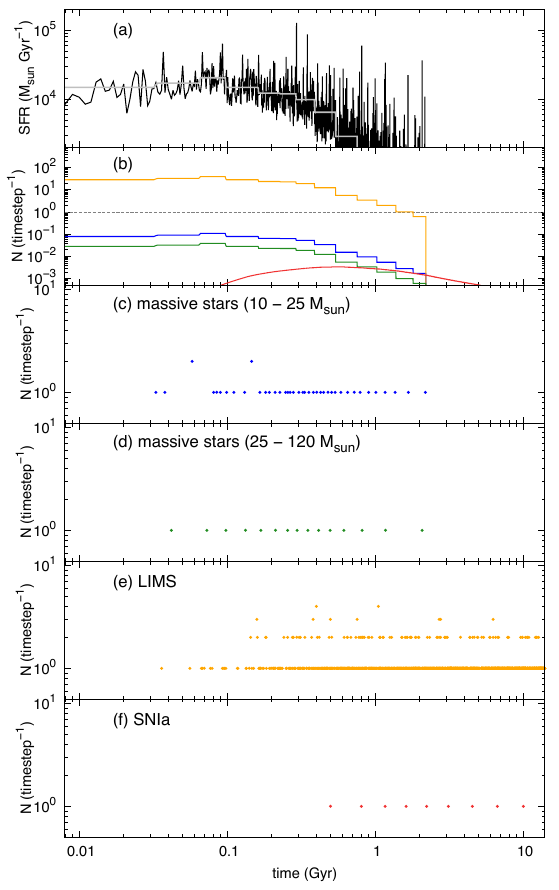}
\caption{Same as Fig.~\ref{fig:nombre}, but the case of hypothetical uFd i is shown
as an example. 
In panel (a), the grey curve shows the star formation rate based on the
observationally-derived star formation history \citep[][see text]{2021ApJ...920L..19S}.
The black curve shows the modelled star formation rate after the rariry is introduced.
Since no r-process event occurs in this uFd, the time variation of 
the number of stars and SNIa is shown.
The total number of stars, SNIa and r-process events throughout the
evolution of each hypothetical uFd is summarized in Table~\ref{tab:nom-uFd}.
}
\label{fig:nom-uFd}
\end{figure}

Figure~\ref{fig:nom-uFd} shows the time variation of the star formation rate
(panel a), 
the number of each astrophysical source (panel b) and the number of dying stars 
and SNIa in a time-step (panels c--f) in uFd i as an example.
The number of massive stars and SNIa is smaller than unity 
(Fig.~\ref{fig:nom-uFd}b) and
the concept of rarity is introduced.

For the calculations, the assumptions described in the main text are 
included in the model. 
We note that physical processes that remove the interstellar gas 
from dwarf galaxies are under discussion. 
While our model represents the simplest cases, it is practical to see 
the impact of rare events on the chemical abundance.

With regard to the rates of the gas accretion and the outflow, 
observed metallicity distributions of individual uFds are 
not always consistent, 
probably due to the small sample and different analyses
\citep[e.g.][for the case of Reticulum~II]{2015ApJ...808...95S,2015ApJ...811...62K,2016ApJ...830...93J}.
We set the efficiency of the outflow $k$ to be 
$k = 8.0\times10^3$ as the median of 
the predicted metallicity distribution is roughly consistent 
with the average metallicity of galaxies of 
${\rm M_* = 5\times10^3~M_{\odot}}$ 
inferred from the stellar mass -- stellar metallicity relation
\citep[][]{2013ApJ...779..102K}. 
Metallicity distributions 
depend on the properties of individual stars, and thus
the efficiency of the outflow adopted here should be regarded 
as a tentative value. 
Also, we roughly set the time-scale of the gas accretion to be 
${\rm \alpha = 3.5~Gyr^{-1}}$. 
The rate of the gas accretion is assumed to become zero 
until the star formation stops ($t \sim 2.2$~Gyr).
As for the yields for the r-process sites, 
the relative contribution is tentatively set to be ${\rm A_r = 0.3}$ 
by adopting the value for Sculptor and Sextans dSphs.

\begin{figure*}
\centering
\includegraphics[scale=1.1]{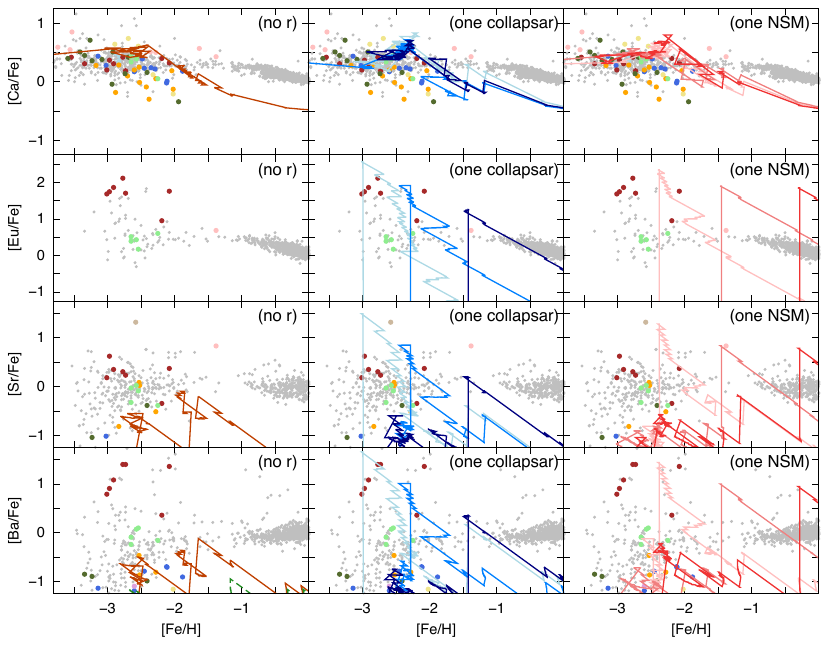}
\caption{The evolution of the [Ca/Fe], [Eu/Fe], [Sr/Fe] and [Ba/Fe] 
ratios of uFds i to vii as a function of metallicity.
The curves are the model predictions for each hypothetical uFd
(see below).
Coloured dots are the abundance ratios of stars in uFds gathered from
the literature.
The colour of the dots, the corresponding uFd and the references
are summarized in Table~\ref{tab:ref-uFd}.
When a single star has been analysed in different studies,
the abundance measured from the data of a higher resolution or 
a larger signal-to-noise ratio is included.
Grey dots are the abundance ratios of stars in the Milky Way.
The references are described in captions of Fig.~\ref{fig:app-h3} for [Ca/Fe],
Fig.~\ref{fig:app-h4} for [Sr/Fe] and Fig.~\ref{fig:r-yield} for [Eu/Fe], 
respectively.
The observational data of the [Ba/Fe] ratio of the Galactic stars are 
taken from \citet{2005A&A...439..129B,2013A&A...552A.128M,2014A&A...562A..71B,2021A&A...649A.126T}.
(left column) The case of no r-process event (uFd i).
Orange and green curves are the abundance ratios predicted by models 
E and F for uFd i, respectively.
Since Eu is mainly produced by the r-process, the [Eu/Fe] ratio of this uFd is
always [Eu/Fe] $\lesssim -1.5$.
(middle column) The case of uFds with one collapsar.
In order of the thickness of the colour (from light to dark), the curves
show the abundance ratios predicted by model A 
for uFds ii, iii and iv, respectively.
(right column) The case of uFds with one NSM.
Similar to the middle column, 
in order of the thickness of the colour, the curves
show the abundance ratios predicted by model A 
for uFds v, vi and vii, respectively.
}
\label{fig:ratio-uFd}
\end{figure*}

Figure~\ref{fig:ratio-uFd} shows the evolution of abundance ratios 
of each hypothetical uFd as a function of metallicity. 
The abundance ratios of stars in uFds are also shown 
for reference. 
The chemical abundance has been gathered from the literature and
the dispersion may partly come from the difference in the analyses.
We recall that the model predictions are for hypothetical uFds with 
the average star formation history, 
and thus the model does not necessarily reproduce the abundance ratios 
of the stars in individual uFds.

\begin{table*}
	\centering
        {\footnotesize
	\caption{UFds and references to the data included in Figure~\ref{fig:ratio-uFd}.}
	\label{tab:ref-uFd}
	\begin{tabular}{lll} 
		\hline
		galaxy & colour & references  \\
		\hline
		Bo\"{o}tes~I  & blue  & \citet{2009AA...508L...1F,2014AA...562A.146I,2016ApJ...826..110F,2023MNRAS.519.1349W}  \\
		Bo\"{o}tes~II & light blue & \citet{2016ApJ...817...41J} \\
		Canes Venatici~II & beige & \citet{2016AA...588A...7F} \\
		Coma Berenices & khaki & \citet{2010ApJ...708..560F,2023MNRAS.519.1349W} \\
		Hercules & orange & \citet{2008ApJ...688L..13K,2011AA...525A.153A,2016AA...588A...7F} \\
                Pieces~II & plum & \citet{2018AA...617A..56S} \\
                Reticulum~II & brown & \citet{2016ApJ...830...93J} \\
		Segue~1 & pink & \citet{2014ApJ...786...74F}  \\
		Tucana~II & dark green & \citet{2018ApJ...857...74C,2023AJ....165...55C}  \\
		Tucana~III & light green & \citet{2017ApJ...838...44H,2019ApJ...882..177M}  \\
		\hline
	\end{tabular}
}
\end{table*}

Mass, metallicity and initial rotating velocity of 
individual stars formed in each uFd are different,
and the evolution of abundance ratios varies among these uFds.
For all of the uFds, the model predicts [Ca/Fe] ratios as high as 
a typical value of the Galactic halo stars ([${\rm \alpha}$/Fe] $\sim~0.4$) 
at [Fe/H] $\lesssim -2.5$, 
and then the ratio decreases with increasing metallicity due to SNe\,Ia. 
The [Ca/Fe] ratio of the hypothetical uFds seems to start to decrease 
at higher metallicity with respect to the uFd stars. 
If stars are efficiently formed at the early time 
(before SNe\,Ia start to contribute to the chemical evolution), 
the metallicity at the ${\rm \alpha}$-knee tends to be high
\citep[e.g.][]{1990ApJ...365..539M}. 
Thus, stars might be more efficiently formed in the hypothetical galaxies 
than in the observed uFds.

Fig.~\ref{fig:ratio-uFd} also shows the impact of rare events 
on the [Eu/Fe], [Sr/Fe] and [Ba/Fe] ratios. 
Since Eu is mainly produced through the r-process, 
the [Eu/Fe] ratio of an uFd without r-process events remains low, 
while the ratio can be high when an r-process event appears. 
In uFd i, no r-process event occurs, 
and the [Eu/Fe] ratio of the galaxy is low ([Eu/Fe] $< -1.5$) 
throughout the evolution. 
Sr and Ba are produced through the s- and the r-processes. 
When r-process events do not occur, the abundance of these elements evolves 
according to the ejection of material from dying stars through the s-process. 
With the presence of rotating massive stars, 
the [Sr/Fe] and [Ba/Fe] ratios of the hypothetical uFds can reach
the values of part of stars of low abundance ratios at [Fe/H] $\lesssim -2$.
Different processes, such as inhomogeneous mixing of gas, can be reflected in the
abundance.
Also, low [Sr/Fe] and [Ba/Fe] ratios of uFds might also be explained by
the weak-r process (\citealt{2006NuPhA.777..676W}, and for the discussion
about the abundance ratios, see e.g. \citealt{2010A&A...524A..58T}).
Thus, it is not conclusive, but
the abundance ratios as low as those of part of the stars in uFds 
might be explained by the ejecta of CCSNe and 
stellar winds of rotating massive stars.
Measurements of the abundance of multiple elements (e.g. Sr, Ba and Eu)
as well as nucleosynthesis calculations on massive stars
are helpful to further discuss astrophysical sources
of the chemical evolution.

In uFds ii, iii and iv (v, vi and vii), the abundance ratios increase 
when a collapsar (NSM) occurs, 
and then the ratios gradually decrease as stars and SNe Ia release iron. 
If the collapsar (NSM) appears at a low metallicity, 
the abundance ratios reach higher values. 
In uFds ii and iii (v), the [Eu/Fe] and [Ba/Fe] ratios can be as high as 
those of stars in Reticulum~II
\citep[see also e.g.][]{2018ApJ...865...87O}.
The abundance ratios of uFd vii seem to significantly increase 
due to the NSM even at a metallicity 
as high as [Fe/H] $\sim~-0.5$. 
In this uFd, the NSM occurs after the star formation has stopped.
When the NSM releases heavy elements in the small amount of 
the interstellar gas, the abundance ratios temporally increase.

In summary, rare events can impact abundance ratios in low-mass galaxies 
at low metallicities as discussed in previous studies. 
Also, a fraction of stars of low abundance ratios in uFds might be 
explained by CCSNe and stellar winds of rotating massive stars. 

\section{The impact of individual events on the chemical abundance in a cosmological context}
\label{sec:app-bb}

Appendix~\ref{sec:app-bb} is available online.




\bsp	
\label{lastpage}
\end{document}